\documentclass[conference]{IEEEtran}
\IEEEoverridecommandlockouts
\usepackage{xspace}

\usepackage{cite}
\usepackage{amsmath,amssymb,amsfonts}
\usepackage{algorithmic}
\usepackage{textcomp}
\def\BibTeX{{\rm B\kern-.05em{\sc i\kern-.025em b}\kern-.08em
    T\kern-.1667em\lower.7ex\hbox{E}\kern-.125emX}}
\usepackage[normalem]{ulem}
\usepackage{multirow}
\usepackage{graphicx, caption, subcaption}
\usepackage{tabularx}

\usepackage{soul} 
\usepackage{xcolor}
\usepackage[hidelinks]{hyperref}
\hypersetup{%
  colorlinks=false,
}
\usepackage{diagbox}
\usepackage[linesnumbered,ruled,vlined]{algorithm2e}
 
\usepackage{listings}

\usepackage{tikz}

\usepackage{enumitem}
\newenvironment{myitemize}
{ \begin{itemize}[leftmargin=0.2in]	
		\vspace{-1ex}	
		\setlength{\itemsep}{0pt}
		\setlength{\parskip}{0pt}
		\setlength{\parsep}{0pt}    }
	{ 	 \end{itemize}                    }

\usepackage{array}

\newcolumntype{L}[1]{>{\raggedright\let\newline\\\arraybackslash\hspace{0pt}}m{#1}}
\newcolumntype{C}[1]{>{\centering\let\newline\\\arraybackslash\hspace{0pt}}m{#1}}
\newcolumntype{R}[1]{>{\raggedleft\let\newline\\\arraybackslash\hspace{0pt}}m{#1}}
 
\newtheorem{definition}{\textbf{Definition}}

\newcommand{\algo }{\emph{MOStream}\xspace} 
\newcommand{\dsc }{\emph{DSC}\xspace}

\newcommand{\cef }{cluster evolution\xspace}

\newcommand{\skm}{\textsf{StreamKM++}\xspace}
\newcommand{\birch}{\textsf{BIRCH}\xspace}
\newcommand{\dstream}{\textsf{DStream}\xspace}
\newcommand{\dbstream}{\textsf{DBStream}\xspace}
\newcommand{\denstream}{\textsf{DenStream}\xspace}
\newcommand{\edmstream}{\textsf{EDMStream}\xspace}
\newcommand{\clustream}{\textsf{CluStream}\xspace}
\newcommand{\slkmeans}{\textsf{SL-KMeans}\xspace}
\newcommand{\kmeans}{\textsf{KMeans}\xspace}

\newcommand{\dbscan}{\textsf{DBSCAN}\xspace}

\newcommand{\benne}{\textsf{Benne}\xspace}

\newcommand{\fct}{\emph{FCT}\xspace}
\newcommand{\kdd}{\emph{KDD99}\xspace}
\newcommand{\sensor}{\emph{Sensor}\xspace}
\newcommand{\insect}{\emph{Insects}\xspace}
\newcommand{\eds}{\emph{EDS}\xspace}
\newcommand{\ods}{\emph{ODS}\xspace}
\newcommand{\dimension}{\emph{Dim}\xspace}

\newcommand{\CFT}{\hyperref[subsec:cf-tree]{\texttt{CFT}}\xspace}
\newcommand{\MCs}{\hyperref[subsec:MCs]{\texttt{MCs}}\xspace}
\newcommand{\CoreT}{\hyperref[subsec:CT]{\texttt{CoreT}}\xspace}
\newcommand{\Grids}{\hyperref[subsec:Grids]{\texttt{Grids}}\xspace}
\newcommand{\DPT}{\hyperref[subsec:DPT]{\texttt{DPT}}\xspace}
\newcommand{\MS}{\hyperref[subsec:meyerson_sketch]{\texttt{AMSketch}}\xspace}

\newcommand{\landmark}{\hyperref[subsec:landmark]{\texttt{LandmarkWM}}\xspace}
\newcommand{\sliding}{\hyperref[subsec:sliding]{\texttt{SlidingWM}}\xspace}
\newcommand{\damped}{\hyperref[subsec:damped]{\texttt{DampedWM}}\xspace}

\newcommand{\nod}{\hyperref[sec:outlier_detection]{\texttt{NoOutlierD}}\xspace}
\newcommand{\od}{\hyperref[subsec:detection]{\texttt{OutlierD}}\xspace}
\newcommand{\buffer}{\hyperref[subsec:buffer]{\texttt{OutlierD-B}}\xspace}
\newcommand{\timer}{\hyperref[subsec:timer]{\texttt{OutlierD-T}}\xspace}
\newcommand{\buffertimer}{\hyperref[subsec:buffertimer]{\texttt{OutlierD-BT}}\xspace}

\newcommand{\norefinement}{\hyperref[sec:no_refinement]{\texttt{NoRefine}}\xspace}
\newcommand{\incrementalRefinement}{\hyperref[sec:incremental_refinement]{\texttt{IncrementalRefine}}\xspace}
\newcommand{\oneshotRefinement}{\hyperref[sec:oneshot_refinement]{\texttt{One-shotRefine}}\xspace}

\usepackage{hyphenat}
\newcommand{\compact}{\vspace{-0pt}}

\usepackage[skip=6pt]{caption}
\newcounter{mycounter}
\usepackage{comment}

\usepackage{makecell}

\usepackage{tcolorbox}
\usepackage{marginnote}

\newcommand{\margii}[2]{
\marginnote{
}
}

\newcommand{\workloadTable}{{
\begin{table}[t]
\centering
\tiny
\caption{Characteristics differences of selected workloads. \textmd{Note that the outliers column refers to whether there are outliers in the final clustering results.}}
\label{tab:Workload}
\resizebox{0.5\textwidth}{!}{%
\begin{tabular}{|c|p{0.5cm}|p{0.5cm}|p{0.5cm}|p{0.6cm}|p{0.6cm}|}
\hline
\textbf{Workload}                           & \textbf{Length}       & \textbf{Dim.}     & \textbf{Cluster Num.}     & \textbf{Outliers}             & \textbf{Evolving Freq.}\\ \hline
{\color{blue}\fct}~\cite{Covertype}         & 581012                & 54                & 7                         & False                         & Low\\ \hline
{\color{blue}\kdd}~\cite{KDD99}             & 4898431               & 41                & 23                        & True                          & Low\\ \hline
{\color{blue}\insect}~\cite{Insect}         & 905145                & 33                & 24                        & False                         & Low\\ \hline
{\color{blue}\sensor}~\cite{SensorData}     & 2219803               & 5                 & 55                        & False                         & High\\ \hline
{\color{blue}\eds}~\cite{DenStream:06}      & 245270                & 2                 & 363                       & False                         & Varying\\\hline
{\color{blue}\ods}~\cite{DenStream:06}      & 100000                & 2                 & 90                        & Varying                       & High\\ \hline
{\color{blue}\dimension}~\cite{MOA_Web}     & 500000                & 20$\sim$100           & 50                        & Low                           & Low\\ \hline
\end{tabular}%
}
\end{table}
}}


\begin{document}

\title{MOStream: A Modular and Self-Optimizing Data Stream Clustering Algorithm}

\author{Zhengru Wang, Xin Wang, Shuhao Zhang
\thanks{Zhengru Wang is with Nvidia, China.}
\thanks{Xin Wang is with the Ohio State University, USA.}
\thanks{Shuhao Zhang is with the Nanyang Technological University, Singapore, e-mail: shuhao.zhang@ntu.edu.sg}
\thanks{}}


\maketitle

\begin{abstract} 
Data stream clustering is a critical operation in various real-world applications, ranging from the Internet of Things (IoT) to social media and financial systems. Existing data stream clustering algorithms, while effective to varying extents, often lack the flexibility and self-optimization capabilities needed to adapt to diverse workload characteristics such as outlier, cluster evolution and changing dimensions in data points. These limitations manifest in suboptimal clustering accuracy and computational inefficiency. In this paper, we introduce \algo, a \underline{m}odular and self-\underline{o}ptimizing data stream clustering algorithm designed to dynamically balance clustering accuracy and computational efficiency at runtime. \algo distinguishes itself by its adaptivity, clearly demarcating four pivotal design dimensions: the summarizing data structure, the window model for handling data temporality, the outlier detection mechanism, and the refinement strategy for improving cluster quality. This clear separation facilitates flexible adaptation to varying design choices and enhances its adaptability to a wide array of application contexts. We conduct a rigorous performance evaluation of \algo, employing diverse configurations and benchmarking it against 9 representative data stream clustering algorithms on 4 real-world datasets and 3 synthetic datasets. Our empirical results demonstrate that \algo consistently surpasses competing algorithms in terms of clustering accuracy, processing throughput, and adaptability to varying data stream characteristics. 
\end{abstract}
\compact
\section{Introduction}
\label{sec:intro}
Data stream clustering (\dsc) is a crucial operation in data stream mining, with applications in network intrusion detection~\cite{KDD99}, social network analysis~\cite{EDMStream:17}, weather forecasting~\cite{weather_forecast:2020}, and financial market analysis~\cite{cai2012clustering}. Unlike traditional batch clustering algorithms such as \kmeans~\cite{KMeans:1967,KMeans:1982} and \dbscan~\cite{DBSCAN:1996}, \dsc algorithms dynamically group incoming data based on attribute similarities and adjust clustering results upon receiving the new streaming data. \dsc algorithms must be efficient~\cite{EmpiricalR:2017,Empirical:2018}, as they are used in high-velocity environments requiring real-time decisions. Additionally, these algorithms need to address unique challenges in data stream scenarios, such as \emph{cluster evolution} and \emph{outlier evolution}~\cite{EDMStream:17,concept_evolution:2010,Concept_Evolution:10,Concept_Evolution:16,Concept_Drift_Survey:2018}, which involve the changing distribution of data and the emergence of new outliers over time and the changing dimensions of data points.

The diverse applications and performance requirements of \dsc have led to the development of numerous \dsc algorithms~\cite{BIRCH:96,Clustream:03,EDMStream:17,DenStream:06,DBStream:16,DStream:2007,SL-KMeans:20,StreamKM++:12,TSFDBSCAN:20,DenStream:06,Stream_Clustering_Survey01,Survey13}. \emph{However, these algorithms, while effective to varying extents, often lack the flexibility and self-optimization capabilities needed to adapt to diverse application contexts.} Typically, \dsc algorithms consist of four main components~\cite{wang2023sesame}: the summarizing data structure, the window model, the outlier detection mechanism, and the refinement strategy. Design choices in each component vary to meet specific needs based on application requirements, data characteristics, or performance objectives. For instance, some algorithms are optimized for processing speed~\cite{DStream:2007}, making them ideal for high-velocity data streams, while others prioritize clustering accuracy~\cite{StreamKM++:12,BIRCH:96}, a critical factor in applications requiring precise cluster assignments. The task of selecting an appropriate \dsc algorithm for real-world, diverse workloads becomes complex due to the interdependent nature of these design choices, each with its own set of trade-offs and performance implications~\cite{wang2023sesame}.

\begin{figure}[t]
	\centering
	\begin{minipage}{1\linewidth}
        \centering
		\includegraphics[width=1\linewidth]{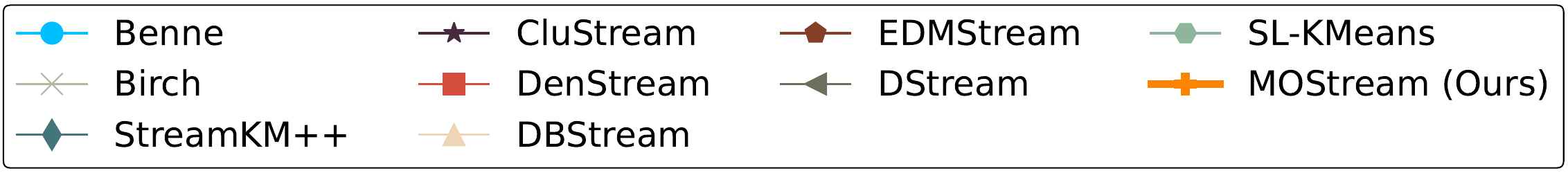}		
	\end{minipage}
	\begin{subfigure}{.245\textwidth}
        \centering
        \includegraphics[width=\linewidth]{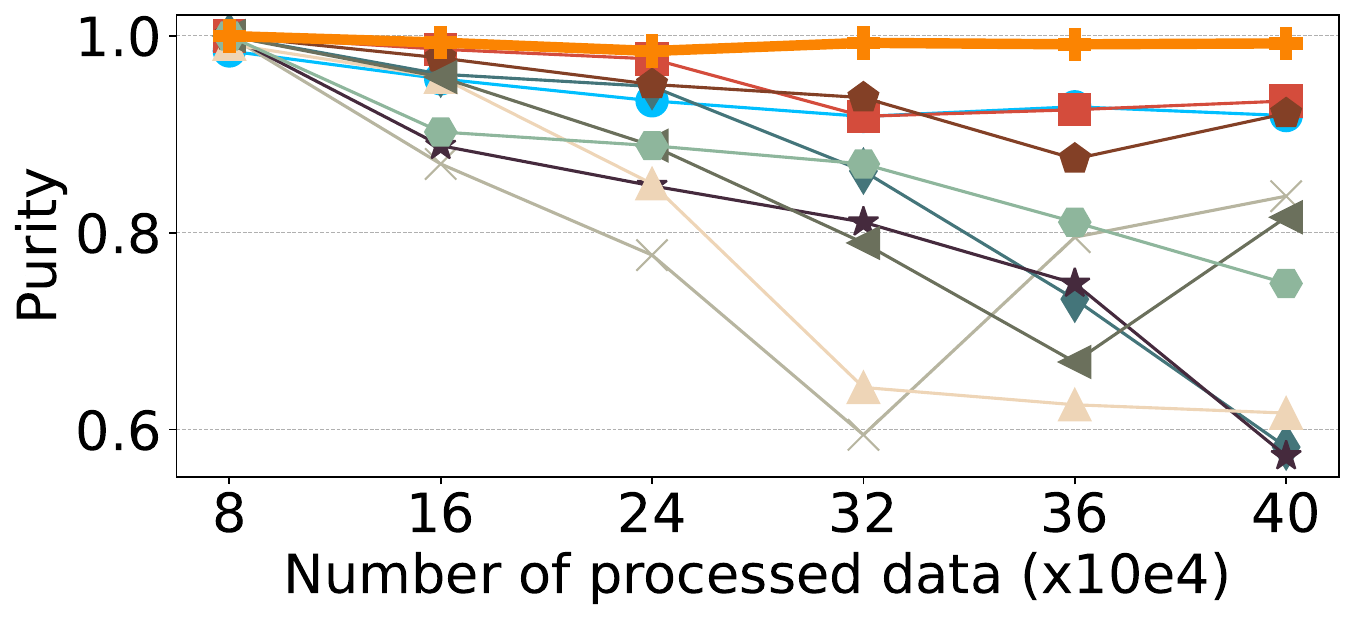}      
         \caption{Accuracy}
         \label{fig:moti_purity}
	\end{subfigure}       
  	\begin{subfigure}{.235\textwidth}
        \centering
        \includegraphics[width=\linewidth]{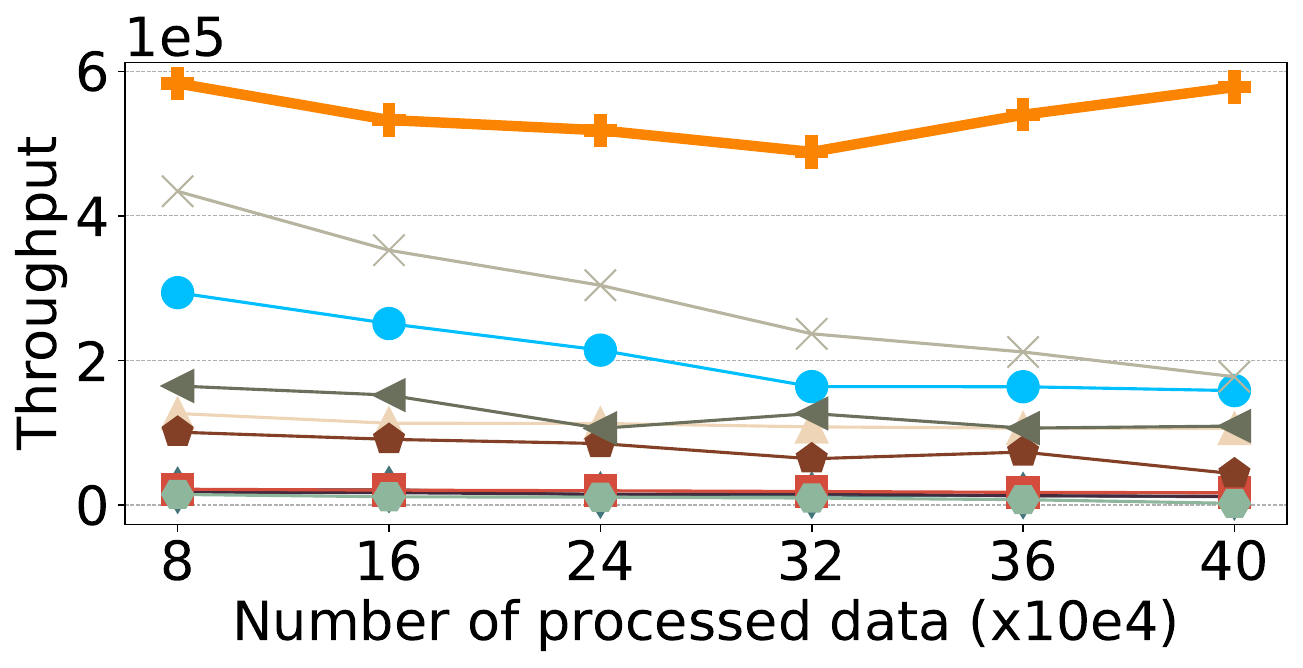}
        \caption{Efficiency}
        \label{fig:moti_throughput}
	\end{subfigure}     
        \caption{Performance comparison of nine representative \dsc algorithms and \algo on \kdd with high frequency of outlier evolution and increasing frequency of cluster evolution in the dataset.}
\label{fig:motivation}
\end{figure}


Figure~\ref{fig:motivation} shows the performance comparison of nine representative \dsc algorithms, including \benne~\cite{wang2023sesame}, \birch~\cite{BIRCH:96}, \skm~\cite{StreamKM++:12}, \clustream~\cite{Clustream:03}, \denstream~\cite{DenStream:06}, \dbstream~\cite{DBStream:16}, \edmstream~\cite{EDMStream:17}, \dstream~\cite{DStream:2007}, and \slkmeans~\cite{SL-KMeans:20} and our proposed algorithm \algo on \kdd with high frequency of outlier evolution and increasing frequency of cluster evolution in the dataset. We observe that both the accuracy and efficiency of existing algorithms suffer from severe degradation when processing more and more streaming data with high frequency of outlier evolution. We attribute this degradation to the inability of their static design choices to dynamically adapt to workload evolution. \benne, for example, allows manual selection of different design options for each of the four design aspects of a \dsc algorithm. While it provides a flexible framework for exploring design choices, it requires manual intervention to select the optimal configuration, which may not be practical in dynamic and rapidly changing data stream environments. In contrast, \algo\footnote{Code is at \url{https://github.com/intellistream/Sesame}} achieves highest accuracy and efficiency during the entire clustering process due to its modular and self-optimizing feature that adapts to changing workload characteristics with little overhead.

Similar to \benne, \algo adopts a \textbf{modular} architecture that allows flexible choices at each design aspect of \dsc algorithms and conducts a  dynamic adjustment of the modular architecture for adapting to the new worload characteristics. By further integrating dynamic adaptability, comprehensive component selection, and novel adaptation techniques, \algo effectively addresses the inherent challenges of \dsc through \textbf{self-optimizing}:

\emph{1. Dynamic Adaptability:} Unlike \benne, which uses a fixed modular configuration, \algo dynamically detects changes in workload characteristics and reconfigures itself in real-time. This capability allows \algo to maintain optimal performance under varying data stream conditions, effectively managing issues such as sudden changes in data dimensionality and outlier frequency.

\emph{2. Comprehensive Component Selection:} \algo includes an automated selection mechanism that not only considers individual component optimization but also evaluates the interplay between different design aspects (summarizing data structure, window model, outlier detection, and refinement strategy). This holistic approach ensures that the algorithm can adapt more comprehensively to diverse data stream scenarios compared to existing methods.

\emph{3. Novel Techniques for Adaptation:} \algo introduces novel techniques for flexible migration and regular stream characteristics detection, which are crucial for maintaining high performance in dynamic environments. These techniques are not present in existing algorithms like \benne and significantly enhance \algo's adaptability and efficiency.

\emph{4. Extensive Empirical Evaluations:} Through extensive empirical evaluations, \algo demonstrates superior performance in both accuracy and efficiency metrics across a wider range of datasets, including real-world and synthetic datasets with evolving characteristics. In-depth studies are also conducted to better comprehend \algo's dynamic composition capability. This breadth of evaluation underscores \algo's robustness and versatility.

Comprehensive analysis and rigorous performance evaluation with 9 representative \dsc algorithms on 4 real-world datasets and 3 synthetic datasets show that \algo consistently surpasses competing algorithms in clustering accuracy, processing throughput, and adaptability to varying data stream characteristics, including varying cluster evolution, outlier evolution, and workload dimensionality. 

The rest of the paper is organised as follows: Section~\ref{sec:background} introduces the background of \dsc algorithms. Section~\ref{sec:algorithm_design} discusses the algorithmic design of \algo. Section~\ref{sec:evaluation} presents our empirical evaluation of \algo. Finally, Section~\ref{sec:relatedwork} reviews additional related work, and Section~\ref{sec:conclusion} concludes the paper.
\compact
\section{Background}
\label{sec:background}

\subsection{Data Stream Clustering Algorithms}
\label{sec:foundational_concepts}
\begin{definition}
A \textbf{data stream} is represented as a sequence of tuples, denoted as \( S = (x_1, x_2, \ldots, x_t, \ldots) \), where \( x_t \) denotes the \( t \)-th data point arriving at time \( t \). Let \( x_t \) be a data point in the stream, and \( D(x_t, C_t) \) be the distance from \( x_t \) to its closest cluster in \( C_t \). If \( D(x_t, C_t) > \delta \), where \( \delta \) is a threshold, \( x_t \) can be considered an \textbf{outlier}. 
\end{definition}

Data stream clustering (\dsc) algorithms efficiently update \( C_t \) to \( C_{t+1} \) in response to changes in data distribution. Unlike traditional clustering algorithms such as \kmeans or \dbscan, which need to recalculate the entire clustering results from scratch upon receiving new data, \dsc algorithms incrementally update their results to ensure more efficient clustering under the data stream scenario. Additionally, \dsc algorithms need to dynamically and promptly handle the evolution of workload characteristics in data stream scenarios. Critical evolution aspects include \emph{cluster evolution}, which refers to the changes of current clusters, such as splitting, merging, disappearing, and emerging upon receiving new streaming data, \emph{outlier evolution}, which refers to the changes in the number of outliers in the data stream, and changes in data dimensionality.

\subsection{Design Aspects of DSC Algorithms}
\dsc algorithms typically consist of several key components to address the challenges posed by data streams. The pros and cons of each design choice are summarized in Table~\ref{tab:design_comparison}.

\begin{table*}[t]
\caption{Design Options Considered in \algo}
\label{tab:design_comparison}
\centering
\begin{tabular}{|p{2.1cm}|p{2.7cm}|p{5.6cm}|p{5.7cm}|}
\hline
\textbf{Aspect} & \textbf{Option} & \textbf{Pros} & \textbf{Cons} \\
\hline
\multirow{5}{*}{Data Structure} & \CFT & Supports a range of operations & - \\
\cline{2-4}
& \CoreT & Can handle dense data streams & Involves tree rebuild \\
\cline{2-4}
& \DPT & Can handle evolving data streams & - \\
\cline{2-4}
& \MCs & Reduces computation load & - \\
\cline{2-4}
& \Grids & Computationally efficient & Limited accuracy with changing data streams \\
\cline{2-4}
& \MS & Ideal for known number of clusters & Involves sketch reconstruction \\
\hline
\multirow{3}{*}{Window Model} & \landmark & Capable of detecting drifts & Sensitive to landmark spacing \\
\cline{2-4}
& \sliding & Responsive to trends & Accuracy loss with small windows \\
\cline{2-4}
& \damped & Smooth adaptation to data evolution & Sensitive to outlier effects \\
\hline
\multirow{4}{*}{Outlier Detection} & \nod & - & May reduce clustering quality \\
\cline{2-4}
& \od & Helps improve accuracy & - \\
\cline{2-4}
& \buffer & Prevents immediate incorporation of outliers & - \\
\cline{2-4}
& \timer & Robust to outliers & Algorithm complexity \\
\cline{2-4}
& \buffertimer & High accuracy & Requires buffer time \\
\hline
\multirow{3}{*}{\shortstack{Refinement \\ Strategy}} & \norefinement & Saves computational resources & Less accurate with evolving data \\
\cline{2-4}
& \oneshotRefinement & Balances efficiency and accuracy & - \\
\cline{2-4}
& \incrementalRefinement & Adapts to new data & Computation overhead \\
\hline
\end{tabular}
\end{table*}

\textit{a) Summarizing Data Structure} provides a compact representation of data points, capturing essential information while minimizing memory usage, as storing the entire data stream is impractical. We consider six different design options for summarizing data structures based on their unique advantages in various scenarios. The \emph{Clustering Feature Tree (\CFT)}~\cite{zhang1997birch} was chosen for its hierarchical clustering capability and efficiency in handling large datasets. The \emph{Coreset Tree (\CoreT)}~\cite{StreamKM++:12} was selected for its strong theoretical guarantees in providing accurate summaries for k-means clustering. The \emph{Dependency Tree (\DPT)}~\cite{EDMStream:17} was preferred for its ability to capture dependency structures among data points, which is particularly beneficial in high-dimensional data. The \emph{Micro Clusters (\MCs)}~\cite{Clustream:03} were utilized for their effective incremental clustering and noise tolerance. The \emph{Grids (\Grids)}~\cite{DStream:2007} were adopted for their straightforward and efficient grid-based clustering approach. Lastly, the \emph{Augmented Meyerson Sketch (\MS)}~\cite{SL-KMeans:20} was included for its capability to provide robust clustering summaries with low computational overhead.

\textit{b) Window Model} handles the temporal aspect of data streams, focusing on recent data points and discarding outdated information to improve the algorithm's clustering capability. We implemented three window models to address different temporal requirements. The \emph{Landmark Window Model (\landmark)}~\cite{StreamKM++:12} is suitable for scenarios where data needs to be segmented based on significant events or time landmarks. The \emph{Sliding Window Model (\sliding)}~\cite{SL-KMeans:20} is effective in continuously updating the clustering model by retaining a fixed-size window of the most recent data. The \emph{Damped Window Model (\damped)}~\cite{DenStream:06, DBStream:16, DStream:2007} is ideal for gradually reducing the influence of older data, which is beneficial in environments with smooth and continuous data evolution.

\textit{c) Outlier Detection Mechanism} identifies and handles noise and outliers in the data stream, preventing them from affecting clustering quality. Detecting outliers in data streams is challenging, particularly with evolving outliers. We examined five categories of outlier detection mechanisms. \emph{Basic Outlier Detection (\od)}~\cite{BIRCH:96} is simple and effective for initial noise filtering. \emph{Outlier Detection with Buffer (\buffer)}~\cite{Clustream:03} enhances basic detection by temporarily storing potential outliers for re-evaluation. \emph{Outlier Detection with Timer (\timer)}~\cite{DStream:2007, DBStream:16} adds a temporal aspect to outlier detection, useful in dynamically evolving data streams. \emph{Outlier Detection with Buffer and Timer (\buffertimer)}~\cite{DenStream:06, EDMStream:17} combines the strengths of buffering and temporal evaluation, ensuring robust outlier management.

\textit{d) Refinement Strategy} updates and refines the clustering model as new data points arrive, adapting to changes in the data distribution. We evaluated three refinement strategies to ensure optimal performance. \emph{No Refinement (\norefinement)}~\cite{BIRCH:96, DStream:2007, EDMStream:17, SL-KMeans:20} is suitable for scenarios where the initial clustering is sufficient, and refinement is unnecessary. \emph{One-shot Refinement (\oneshotRefinement)}~\cite{Clustream:03, DenStream:06, StreamKM++:12, DBStream:16} is effective for scenarios requiring a single comprehensive update after initial clustering. \emph{Incremental Refinement (\incrementalRefinement)} periodically applies one-shot refinement during the online clustering process, ensuring continuous adaptation to data stream changes.

\compact
\section{Design of \algo}
\label{sec:algorithm_design}
In this section, we first discuss our motivation towards the design of \algo in Section~\ref{subsec:motivation}. We then provide a brief summary of the general workflow of \algo in Section~\ref{subsec:workflow}. After that, we discuss the three key designs of \algo to self-adjust its design components in response to dynamically detected workload characteristics of the data stream, including Regular Stream Characteristics Detection in Section~\ref{subsec:auto-selection}, Automatic Design Choice Selection in Section~\ref{subsec:regular-detection}, Flexible Algorithm Migration in Section~\ref{subsec:migration}.

\subsection{Motivation} 
\label{subsec:motivation} 
Despite numerous \dsc algorithms, all commonly incorporate the four key design components mentioned earlier. As summarized in Table~\ref{tab:design_comparison}, Wang et al.~\cite{wang2023sesame} revealed that no single design choice consistently guarantees good performance across varying workload characteristics and optimization targets. Based on these findings, Wang et al. proposed \benne, the first \dsc algorithm with a modular architecture that selects different design options from four design aspects. While \benne achieved state-of-the-art performance, it struggles with changes in stream characteristics where a fixed modular composition may not be effective. For instance, when the frequency of cluster evolution of the data stream suddenly increases, using the old configuration designed for low cluster evolution frequency can result in poor performance. To address this issue, we designed \algo, a modular and self-optimizing data stream clustering algorithm. Unlike \benne, \algo dynamically detects changes in workload characteristics and reconfigures itself with the best modular composition.

\subsection{General Workflow}
\label{subsec:workflow}
The \algo's adaptability is rooted in the flexible configuration of four core components: the summarizing data structure (\textit{struc.}), window model (\textit{win.}), outlier detection mechanism (\textit{out.}), and refinement strategy (\textit{ref.}). \algo consists of four main functions to integrate these core components together. Specifically,
\emph{Insert Fun.(...)} adds the new streaming data to the current summarizing data structure, \emph{Window Fun.(...)} manages the data points in the summarizing data structure, \emph{Outlier Fun.(...)} assesses if the current input point qualifies as an outlier. \emph{Refine Fun.(...)} runs the offline clustering before obtaining the final results.

Guided by real-time stream characteristics detected from a sample queue (\textit{queue}), predefined threshold values (\textit{thresholds}), and a performance objective (\textit{o}), \algo dynamically tailors its components to meet the specialized requirements of diverse applications and data streams. In particular, the Regular-detection Fun (Section~\ref{subsec:regular-detection}) detects the changes of characteristics in the current data stream, Auto-selection Fun (Section~\ref{subsec:auto-selection}) selects the best modular configurations based on the detected changes of characteristics, Flexible-migration Fun (Section~\ref{subsec:migration}) migrates the clustering results stored in the old modular configuration to the new one. 

\begin{algorithm}[t]
\scriptsize
\caption{Execution flow of \algo.}
\label{alg:algo}
    \KwData{$p$ \tcp{Input point}}
    \KwData{$c$ \tcp{Temporal Clusters}}
    \KwData{$struc.$ \tcp{Type of summarizing data structure}}   
    \KwData{$win.$ \tcp{Type of window model}}
    \KwData{$out.$ \tcp{Type of outlier detection mechanism}}
    \KwData{$ref.$ \tcp{Type of refinement strategy}}
    \KwData{$queue$ \tcp{Batch of stream for detection}}
    \KwData{$thresholds$ \tcp{Identifying characteristics' changes}}
    \KwData{$o$ \tcp{User's primary performance objective}}
    \tcp{Online Phase}
    Initializing parameters and design choices \;
    \While{!stop processing of input streams}{
      \eIf{\text{$queue$ is full}}{
            $struc\_{old}.$ = $struc.$ \;
            $characteristics$ = \emph{\textcolor{blue}{Regular-detection Fun.(...)}} \;
            $struc.$, $win.$, $out.$, $ref.$ = \emph{\textcolor{blue}{Auto-selection Fun.(...)}} \;
            \emph{\textcolor{blue}{Flexible-migration Fun.(...)}}\;
            Empty $queue$ \;
        }{
            $queue$.push($p$) \;
        } 
        \emph{\textcolor{blue}{Window Fun. (...)}}\;
        \eIf{\text{$out.$ != \nod}}{
            $b$ $\gets$ \emph{\textcolor{blue}{Outlier Fun. (...)}}\;
            \If{\text{$b$ = false}}{
                \emph{\textcolor{blue}{Insert Fun. (...)}}\tcp{Insert $p$ to $s$ and update $s$}            
            }
        }{
            \emph{\textcolor{blue}{Insert Fun. (...)}}\tcp{Insert $p$ to $s$ and update $s$}
        }
        \If{\text{$ref.$ = \incrementalRefinement}}{
            \emph{\textcolor{blue}{Refine Fun. ($ref.$)}};
        }
    }
    \tcp{Offline Phase}
    \If{\text{$ref.$ = \oneshotRefinement}}{
        \emph{\textcolor{blue}{Refine Fun. ($ref.$)}};
    }
\end{algorithm}

Algorithm~\ref{alg:algo} outlines the high-level execution flow of \algo. To leverage advantageous design options under different workload characteristics and optimization targets, \algo operates using a bifurcated execution strategy, comprising an online phase and an offline phase.
In the online phase, the algorithm sequentially processes incoming data points. Initially, parameters and design choices for the summarizing data structure (\textit{struc.}), window model (\textit{win.}), outlier detection mechanism (\textit{out.}), and refinement strategy (\textit{ref.}) are initialized. For each incoming data point \( p \), it is pushed into the sample queue (\textit{queue}). When the queue is full, the algorithm performs regular detection to assess the stream characteristics. The current design choices are stored as \textit{struc\_old.}, and the characteristics are detected using the \emph{Regular-detection Function}. Based on the detected characteristics, the \emph{Auto-selection Function} dynamically selects the optimal design choices for \textit{struc.}, \textit{win.}, \textit{out.}, and \textit{ref.}. The \emph{Flexible-migration Function} then facilitates the transition to the new design choices, and the queue is emptied after the selection process. The \emph{Window Function} processes the data points within the current window model. If an outlier detection mechanism is specified (\textit{out.} $\neq$ \nod), the algorithm uses the \emph{Outlier Function} to check if the point \( p \) is an outlier. If \( p \) is not an outlier, it is inserted into the summarizing data structure and the structure is updated using the \emph{Insert Function}. If no outlier detection is specified, the point \( p \) is directly inserted into the summarizing data structure. If the refinement strategy is set to \textit{Incremental Refinement}, the algorithm updates the clustering model using the \emph{Refine Function}. In the offline phase, after processing the entire data stream, if the refinement strategy is set to \textit{One-shot Refinement}, the clustering model is updated using the \emph{Refine Function} to finalize the clustering results.

\subsection{Regular Stream Characteristics Detection}
\label{subsec:regular-detection}
Various stream characteristics such as data dimensionality, cluster evolution, and the number of outliers are subject to change in evolving data streams. To make informed decisions about the most appropriate design choices for real-time clustering, \algo must first ascertain the current characteristics of the data stream. Algorithm~\ref{algo:auto_detection} delineates the procedure \algo employs to automatically detect key attributes like dimensionality, cluster evolution, and the number of outliers in the current data stream.

\begin{algorithm}[t]
\scriptsize
\caption{Regular-detection Fun. of \algo}
\label{algo:auto_detection}
\SetKwFunction{FMain}{Regular-detection Fun.}
\SetKwProg{Fn}{Function}{:}{}
\tcc{$queue$: a queue of new input stream data stored for detecting stream characteristic changes, $centers$: previous clustering centers of the stream data, $thresholds$: threshold values for identifying the characteristic}
\Fn{\FMain{$queue, centers, threshold$}}{
    \tcc{Initialize stream characteristics}
    \text{$characteristics$ = null}\;
    \text{$high\_dim\_data$ = $var$ = $num\_of\_outliers$= 0}\;
    \text{$new\_center$ = mean($queue$)}\;
    \For{$s \in queue$}{
        \If{$s.dim  > T_d$}{
            $high\_dim\_data$ += 1 \;
        }
        $var$ = UpdateVariance($var, s, new\_centers$)\;
        \If{$min(dist(s, centers))  > thresholds.dist$}{
            $num\_of\_outliers$ += 1 \;
        }
    }
    \eIf{$high\_dim\_data > num(samples) / 2$}{
        $characteristics.high\_dimension = true$ \;
    }{
        $characteristics.high\_dimension = false$ \;
    }
    \eIf{$var > thresholds.variance$}{
        $characteristics.frequent\_evolution = true$ \;
    }{
        $characteristics.frequent\_evolution = false$ \;
    }
    \eIf{$num\_of\_outliers > num(samples) / 2$}{
        $characteristics.many\_outliers = true$ \;
    }{
        $characteristics.many\_outliers = false$ \;
    }
    Return $characteristics$; 
}
\end{algorithm}

Specifically, the algorithm initializes various counters and variables to store stream characteristics at Lines 3-5 of Algorithm~\ref{algo:auto_detection}. For each new data point in the $queue$ (Line 6), \algo evaluates its dimensionality. If the dimension exceeds the threshold $T_d$, the counter $high\_dim\_data$ is incremented  (Line 7). The variance of the data stream is updated at Line 8. The algorithm also checks whether each data point is an outlier based on its distance to the current clustering centers (Line 9-10). 
After processing all the data points in the $queue$, \algo sets the $characteristics.high\_dimension$ attribute to ``true'' or ``false'' based on the value of $high\_dim\_data$ (Lines 11-14). Similarly, the $characteristics.frequent\_evolution$ attribute is determined based on the calculated variance (Lines 15-19), and the $characteristics.many\_outliers$ attribute is set based on the number of outliers detected (Lines 20-23).

\begin{algorithm}[t]
\scriptsize
\caption{Auto-selection Fun. of \algo}
\label{algo:auto_selection}
\SetKwFunction{FMain}{Auto-Selection Fun.}
\SetKwProg{Fn}{Function}{:}{}
\tcc{$objective$: User's primary performance objective (Accuracy, Efficiency, or Balance); $data$: Input data stream; $characteristics$: Stream characteristics detected by regular-detect Fun.}
\Fn{\FMain{$objective$, $data$, $characteristics$}}{
    \tcc{Initialize  four modular design components}
    \text{$struc.$= $win.$ = $out.$= $ref.$ = null}\;
    \eIf{$objective$ = Accuracy}{
        \tcc{Select modules with high accuracy based on input data stream characteristics}
        \eIf{$characteristics.frequent\_evolution = true$}{
            $struc.$ = \MCs \;
        }{
            $struc.$ = \CFT \;
        }
        \eIf{$characteristics.many\_outliers = true$}{
            $win.$ = \landmark \;
            $out.$ = \buffertimer \;
        }{
            $win.$ = \damped \;
            \eIf{$characteristics.high\_dimension = true$}{
                $out.$ = \buffer \;
            }{
                $out.$ = \buffertimer \;
            }
        }
        $ref.$ = \incrementalRefinement \;
    }{
        \tcc{Choose a structure with high efficiency based on input data stream characteristics}
        \eIf{$characteristics.frequent\_evolution = true$}{
            $struc.$ = \DPT \;
            $win.$ = \landmark \;
        }{
            $struc.$ = \Grids \;
            $win.$ = \sliding \;
        }
        $out.$ = \nod \;
        $ref.$ = \norefinement \;
    }
    
    Return $struc.$, $win.$, $out.$, $ref.$; 
}
\end{algorithm}

\subsection{Automatic Design Choice Selection}
\label{subsec:auto-selection}
Upon receiving the stream characteristics from the \emph{Regular-detection Function}, \algo proceeds to select the most appropriate design choices based on these characteristics and the user-defined optimization objective. The detailed selection process is delineated in Algorithm~\ref{algo:auto_selection}.

If the optimization objective is set to Accuracy (Line 3), \algo selects the summarizing data structure based on the frequency of cluster evolution. Specifically, \MCs is chosen if frequent cluster evolution is detected (Line 6), otherwise \CFT is selected (Line 8). For the window model and outlier detection mechanisms, \algo selects \landmark and \buffertimer if many outliers are present (Lines 10-11). If outliers are scarce, \damped is chosen as the window model, and the choice between \buffer and \buffertimer for outlier detection is influenced by the data's dimensionality (Lines 13-16). The refinement strategy is set to \incrementalRefinement (Line 18). If the optimization objective is set to Efficiency (Line 20), \algo selects either \DPT or \Grids as the summarizing data structure based on the frequency of cluster evolution (Lines 22-25). The window model is also selected accordingly, with \landmark chosen for frequent cluster evolution and \sliding otherwise (Lines 24-25). The outlier detection mechanism is set to \nod (Line 26), and the refinement strategy is set to \norefinement (Line 27).

 
\begin{algorithm}[t]
\scriptsize
\caption{Flexible-migration Fun. of \algo.}
\label{alg:flexible_migration}
\SetKwFunction{FMain}{Flexible-migration Fun.}
\SetKwProg{Fn}{Function}{:}{}
\tcc{$objective$: User's primary performance objective (Accuracy or Efficiency); $struc.$: New selected summarizing data structure; $struc\_{old}.$: Previous selected summarizing data structure; $out.$: Previous selected outlier detection mechanism}
\Fn{\FMain{$objective$, $struc.$, $struc\_{old}.$}}{
    \If{\text{$struc.$ != $struc\_{old}.$}}{
        Extract the clustering centers $c$ from $struc.$\;
        \eIf{$objective$ = Accuracy}{
            \tcc{Transfer the old clustering results to the new summarizing data structures for the accuracy or balance objective}
            \If{$out.$ = \buffer or \buffertimer}{
                Extract the outliers $o$ from $out.$\;
            }
            Initialize the new summarizing data structure $struc.$ with $c$ and $o$\;
        }{
            \tcc{Sink the old clustering results for the efficiency objective}
            Sink $c$ into the output\;
            Create a blank object for initializing  $struc.$\; 
        }
    }
}
\end{algorithm}
\subsection{Flexible Algorithm Migration}
\label{subsec:migration}
Due to the significant structural differences among various summarizing data structures, it is not feasible to directly transfer clustering information from the old summarizing data structure to the new one. To address this, \algo employs a migration function, delineated in Algorithm~\ref{alg:flexible_migration}. 
Specifically, if the newly selected summarizing data structure (\textit{struc.}) differs from the previously selected one (\textit{struc\_old.}) (Line 2), the algorithm adapts as follows:
For the \textbf{Accuracy Objective} (Lines 4-7), \algo extracts the clustering centers (\textit{c}) from the old summarizing data structure (Line 3). Additionally, if the old outlier detection mechanism (\textit{out.}) is either \buffer or \buffertimer, outliers (\textit{o}) are also extracted (Line 6). These centers and outliers are then used to initialize the new summarizing data structure (\textit{struc.}) (Line 7).
For the \textbf{Efficiency Objective} (Lines 8-10), \algo extracts the clustering centers (\textit{c}) from the old summarizing data structure and sinks them into the output to avoid computational overhead (Line 9). A new, empty object is created to initialize the new summarizing data structure (\textit{struc.}) (Line 10).

\subsection{Further Implementation Details} 
\algo is designed as a three-threaded pipeline to process data streams, approximating a realistic computational environment. Inter-thread communication is facilitated through a shared-memory queue, reducing latency associated with network transmissions. The \emph{Data Producer Thread} loads benchmark workloads into memory and enqueues each data point into a shared queue. To simulate high-throughput, the input arrival rate is immediate, eliminating idle time for the algorithm. The \emph{Data Consumer Thread} executes the \dsc algorithm, processing the data stream by dequeuing input tuples from the shared queue and generating temporal clustering results. Efficiency metrics are captured in this thread for consistent comparison of various \dsc algorithms. The \emph{Result Collector Thread} stores the temporal clustering results from the Data Consumer Thread. It computes accuracy metrics to minimize interference with efficiency measurements. Clustering quality is evaluated using purity~\cite{10.1145/1557019.1557115}, and the ability to handle \cef is assessed using CMM~\cite{CMM:2011}.
\compact
\section{Experimental Analysis}
\label{sec:evaluation}
In this section, we present the evaluation results. All experiments are carried out on an Intel Xeon processor. Table~\ref{tab:specification} summarizes the detailed specification of the hardware and software used in our experiments.
\begin{table}[t]
\centering
\caption{Specification of our evaluation platform}
\label{tab:specification}
\resizebox{0.45\textwidth}{!}{%
\begin{tabular}{|l|l|}
\hline
Component               & Description                                \\ \hline
Processor               & Intel(R) Xeon(R) Gold 6338 CPU @ 2.00GHz   \\ \hline
L3 Cache Size           & 48MiB                                      \\ \hline
Memory                  & 1024GB DDR4 RAM                             \\ \hline
OS                      & Ubuntu 22.04.4 LTS                         \\ \hline
Kernel                  & Linux 5.15.0-100-generic                    \\ \hline
Compiler                & GCC 11.4.0 with -O3                        \\ \hline
\end{tabular}%
} 
\end{table}

\workloadTable
\subsection{Experiment Setups}
\label{subsec:benchmarks}
Table~\ref{tab:Workload} provides a summary of the datasets for evaluation. Our dataset selection is governed by two primary criteria. First, we aim for a fair and comprehensive evaluation by including the three most frequently used datasets across the algorithms summarized in Table~\ref{tab:Workload}. Specifically, \fct (Forest CoverType) is employed by algorithms such as \slkmeans, \skm, \edmstream, and \dbstream. The \kdd dataset is utilized by \skm, \dstream, \edmstream, \denstream, \clustream, and \dbstream, while \sensor is specifically used by \dbstream. In addition to these three classical datasets, we incorporate a more recent dataset, \insect, which was proposed in 2020~\cite{Insect}.
Second, although some previous studies have proposed synthetic datasets, these datasets are not publicly available. To address this limitation and to evaluate the algorithm under varying workload characteristics, as delineated in Table~\ref{tab:Workload}, we design three synthetic datasets: \eds, \ods, and \dimension. The \eds dataset contains varying frequencies of the occurrence of \cef, \ods features a time-varying number of outliers at different stages, and \dimension comprises data with extremely high dimensions.
A detailed account of each dataset is as follows:
\begin{myitemize}
    \item \textbf{\fct} (Forest CoverType)~\cite{Covertype} consists of tree observations from four areas of the Roosevelt National Forest in Colorado. It is a high-dimensional dataset with 54 attributes, and each data point has a cluster label indicating its tree type. The dataset contains no outliers.
    \item \textbf{\kdd}~\cite{KDD99} is a large dataset of network intrusion detection stream data collected by the MIT Lincoln Laboratory. It is also high-dimensional and contains a significant number of outliers, making it suitable for testing outlier detection capabilities.
    \item \textbf{\insect}~\cite{Insect} is the most recent dataset, generated by an optical sensor that measures insect flight characteristics. It is specifically designed for testing the clustering of evolving data streams.
    \item \textbf{\sensor}~\cite{SensorData} contains environmental data such as temperature, humidity, light, and voltage, collected from sensors deployed in the Intel Berkeley Research Lab. It is a low-dimensional dataset with only five attributes but has a high frequency of \cef.
    \item \textbf{\eds} is a synthetic dataset used in previous works~\cite{DenStream:06} to study \cef. It is divided into five stages according to evolving frequency, allowing for a comparative analysis of algorithmic performance across these stages.
    \item \textbf{\ods} is another synthetic dataset, distinct from \eds in that its second half is composed entirely of outliers, enabling an analysis of algorithmic performance under varying numbers of outliers.
    \item \textbf{\dimension} is generated using the RandomTreeGenerator from the MOA framework~\cite{MOA_Web}. It features data points with dimensions ranging from 20 to 100 and 50 classes, with other specific configurations set to default.
\end{myitemize}

\textbf{Evaluation Metrics.}
To measure the clustering quality of \dsc algorithms, we use the widely adopted metric purity~\cite{10.1145/1557019.1557115}, which assesses how well the data points within each cluster belong to the same class. Additionally, we employ CMM~\cite{CMM:2011}, specifically designed to evaluate the ability of \dsc algorithms to handle cluster evolution in the stream. To assess performance, we introduce the throughput metric, which represents the amount of data the algorithm can process within a certain period.

\begin{figure}[t]
	\centering
	\begin{minipage}{\linewidth}
		\includegraphics[width=1\linewidth]{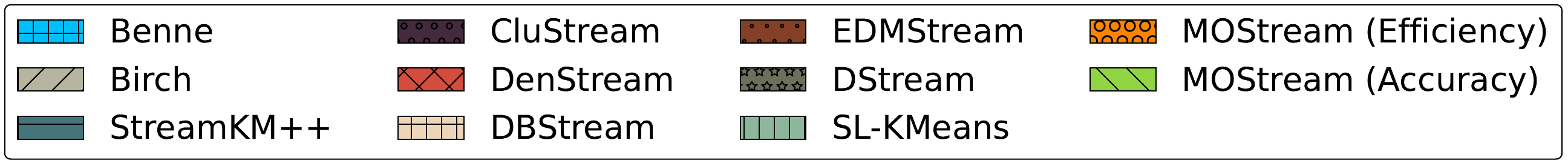}		
	\end{minipage}
	\begin{subfigure}{.24\textwidth}
        \centering
        \includegraphics[width=\linewidth]{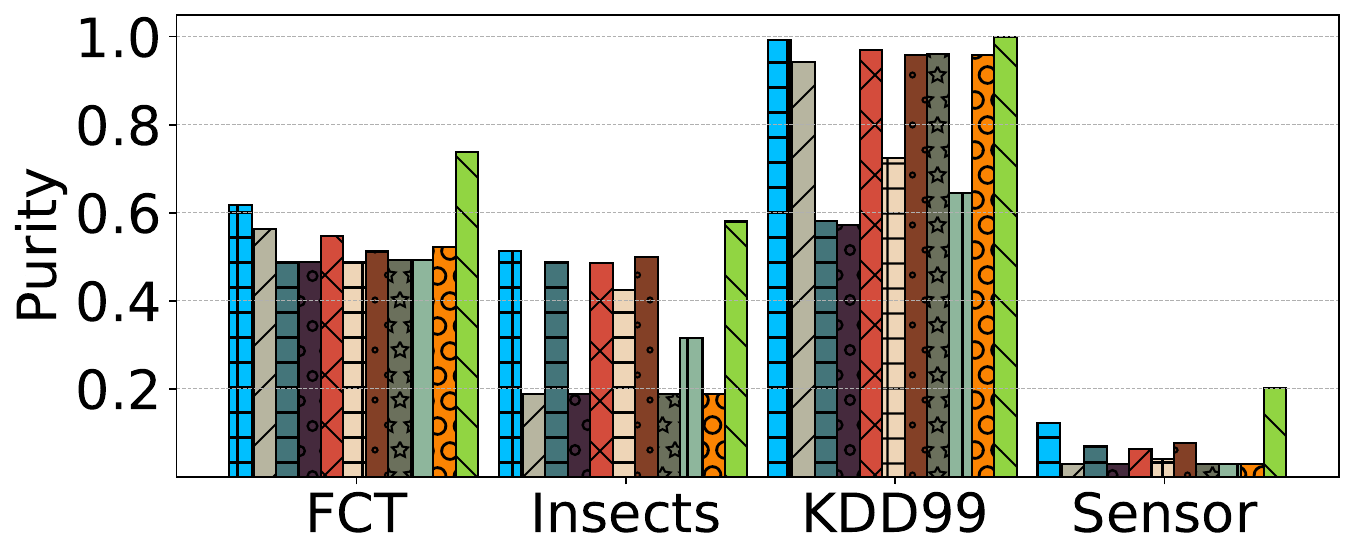}      
         \caption{Accuracy}
         \label{fig:general_purity}
	\end{subfigure}       
  	\begin{subfigure}{.235\textwidth}
        \centering
        \includegraphics[width=\linewidth]{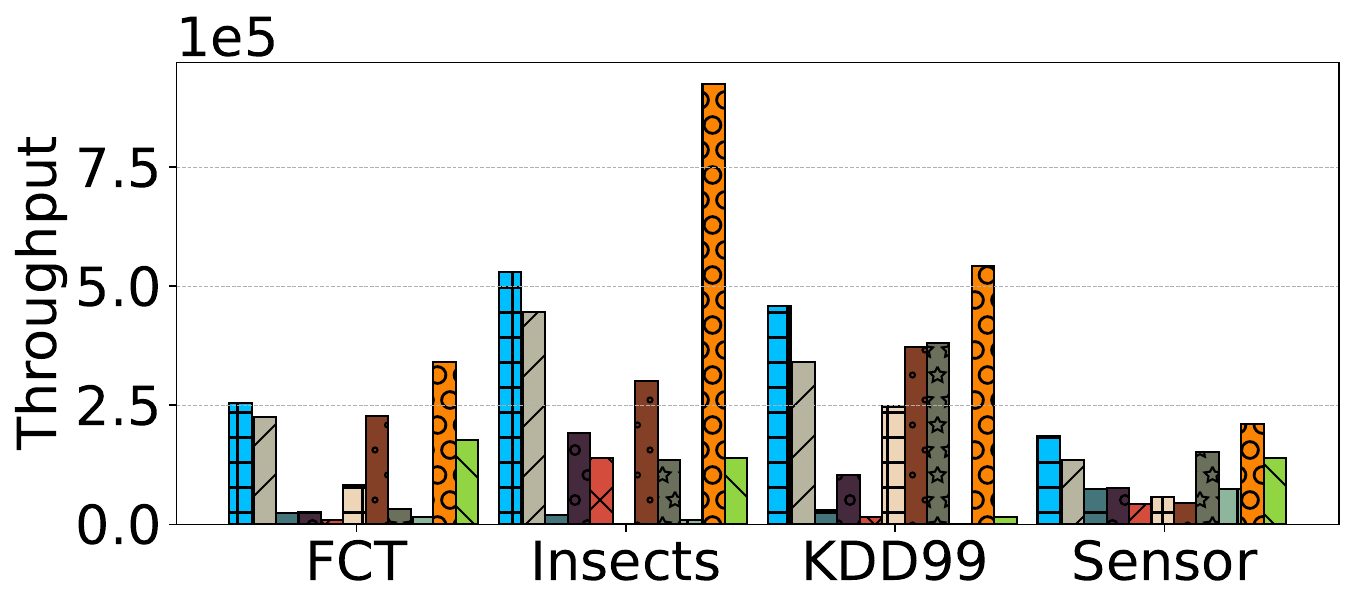}
        \caption{Efficiency}
        \label{fig:general_throughput}
	\end{subfigure}     
    \caption{Performance Comparison on Four Real-world Workloads. The throughput of \dbstream on \insect dataset and the throughput of \slkmeans on \kdd dataset are much lower than other baselines thus neglected to be shown in the figure.}
    \label{fig:general_comparison}
\end{figure}
 
\subsection{General Evaluation of Clustering Behavior}
\label{subsec:static_evaluation}
The versatility of \algo allows it to be configured into two primary variants: \emph{\algo (Accuracy)} and \emph{\algo (Efficiency)}. These variants cater to different optimization objectives based on distinct algorithmic design decisions. We initiate our evaluation by comparing the clustering behavior of these \algo variants with nine existing \dsc algorithms. The evaluation is conducted on four real-world datasets—\fct, \kdd, \insect, and \sensor. The outcomes are illustrated in Figure~\ref{fig:general_comparison}. We can see that \emph{\algo} (Accuracy) attains state-of-the-art purity across all four real-world datasets. Conversely, \emph{\algo (Efficiency)} exhibits purity levels comparable to existing algorithms but surpasses them in throughput. These observations confirm that by judiciously selecting and integrating different design elements, \algo can either optimize for accuracy or efficiency. However, achieving both optimal accuracy and efficiency simultaneously remains elusive, corroborating our analysis regarding the trade-off between these two metrics.

\subsection{Clustering over Varying Workload Characteristics}
\label{subsec:dynamic_evaluation}
\begin{figure}[t]
	\begin{minipage}{1\linewidth}
		\includegraphics[width=1\linewidth]{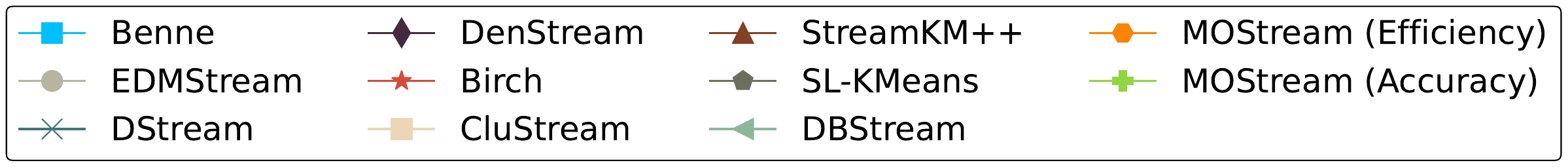}		
	\end{minipage}
 
	\begin{subfigure}{.245\textwidth}
        \centering
        \includegraphics[width=\linewidth]{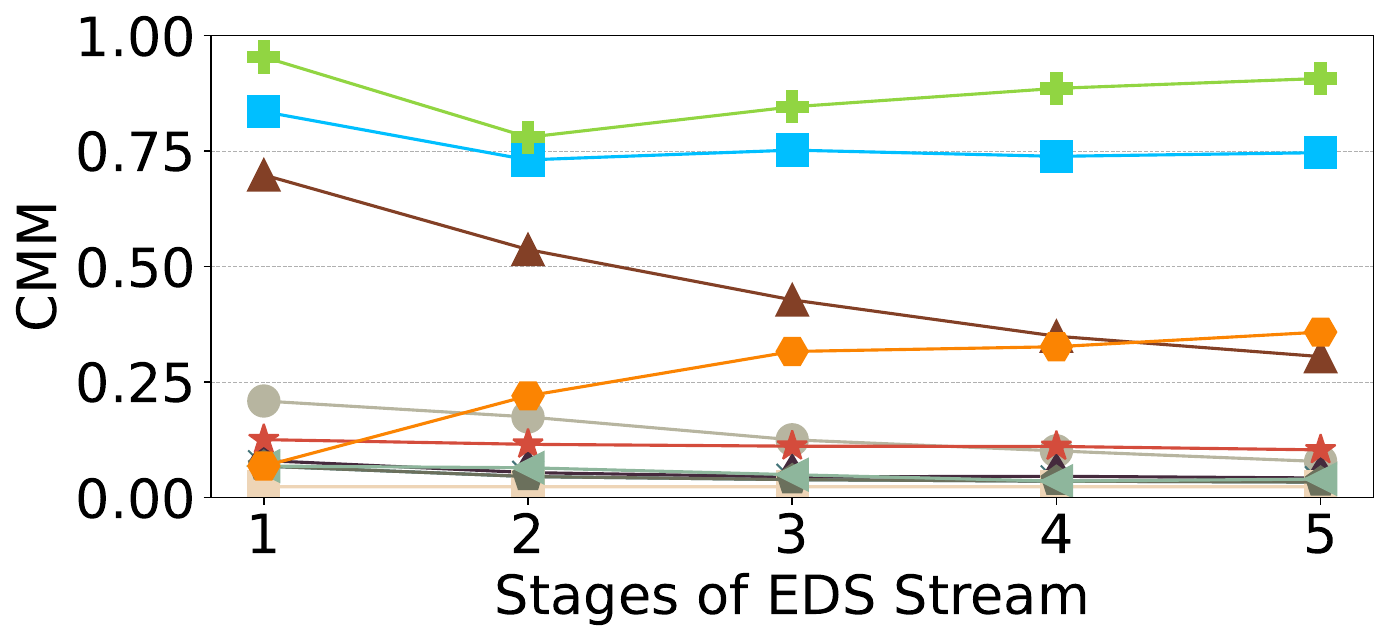}      
         \caption{\eds Accuracy}
         \label{fig:eds_cmm}
	\end{subfigure}       
  	\begin{subfigure}{.23\textwidth}
        \centering
        \includegraphics[width=\linewidth]{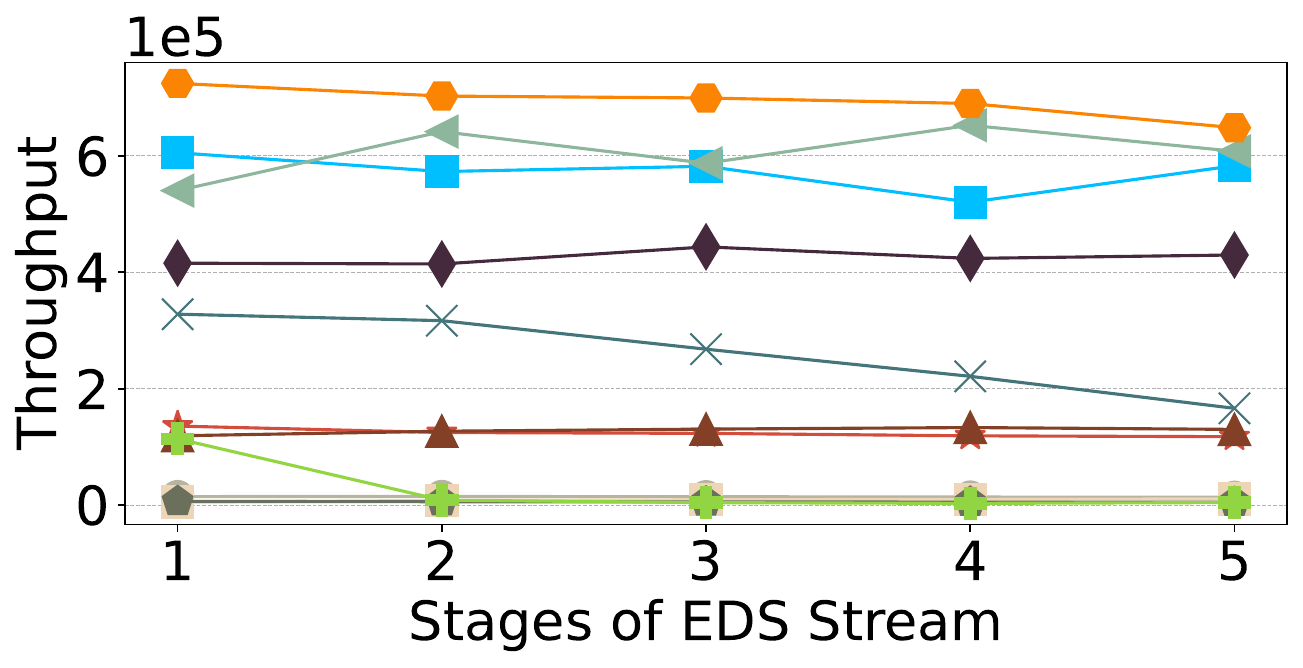}
        \caption{\eds Efficiency}
        \label{fig:eds_throughput}
	\end{subfigure}  
 \begin{subfigure}{.245\textwidth}
        \centering
        \includegraphics[width=\linewidth]{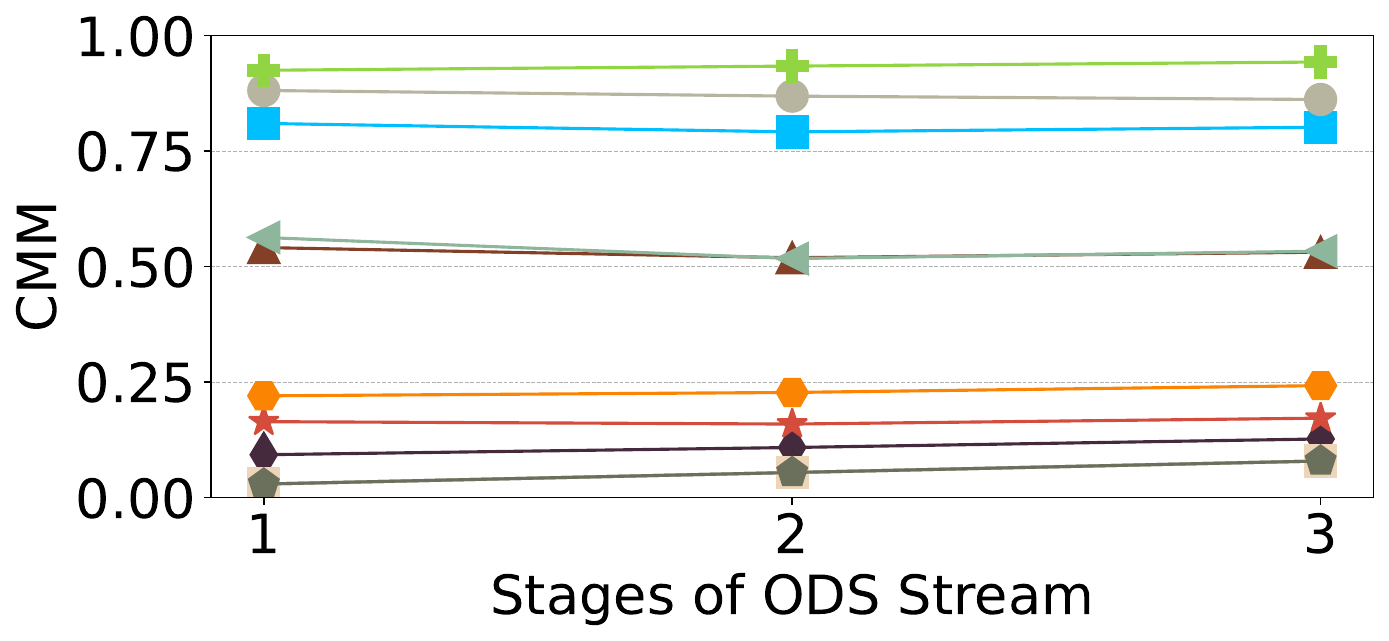}      
         \caption{\ods Accuracy}
         \label{fig:ods_purity}
	\end{subfigure}       
  	\begin{subfigure}{.238\textwidth}
        \centering
        \includegraphics[width=\linewidth]{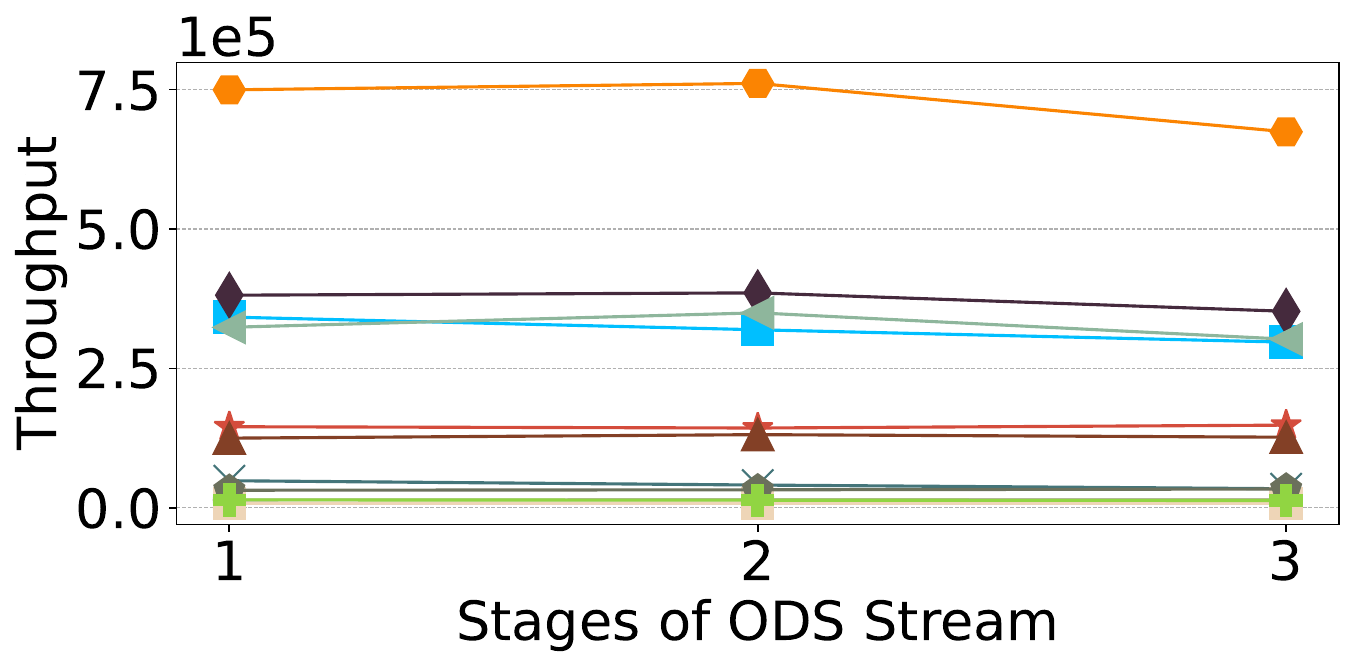}
        \caption{\ods Efficiency}
        \label{fig:ods_throughput}
	\end{subfigure}  
    \caption{Performance Comparison on \eds (Figure (a), (b)) and \ods (Figure (c), (d)) workloads with varying cluster or outlier evolution frequency. \eds is divided into five main stages according to the cluster evolution frequency that increases with the increase of the stages.  \ods is divided into three main stages according to the outlier evolution frequency that increases with the increase of the stages.}
        \label{fig:evolve_comparison}
\end{figure}

\begin{figure}[t]
	\centering
	\begin{minipage}{1\linewidth}
        \centering
		\includegraphics[width=0.7\linewidth]{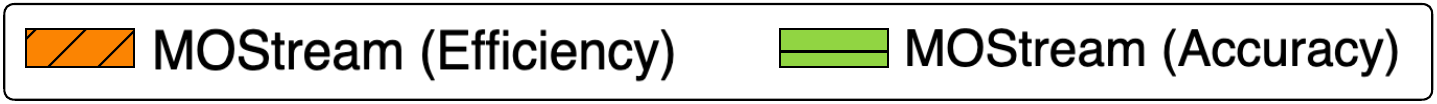}		
	\end{minipage}
	\begin{subfigure}{.245\textwidth}
        \centering
        \includegraphics[width=\linewidth]{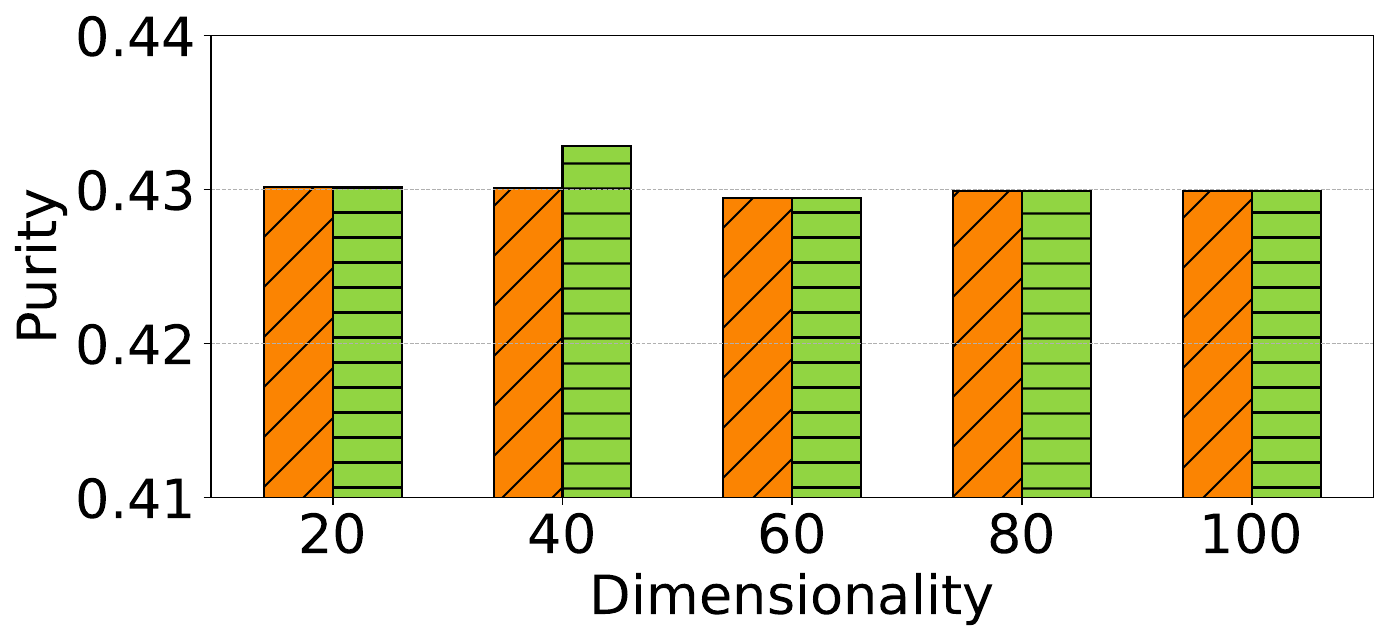}      
         \caption{Accuracy}
         \label{fig:dim_purity}
	\end{subfigure}       
  	\begin{subfigure}{.235\textwidth}
        \centering
        \includegraphics[width=\linewidth]{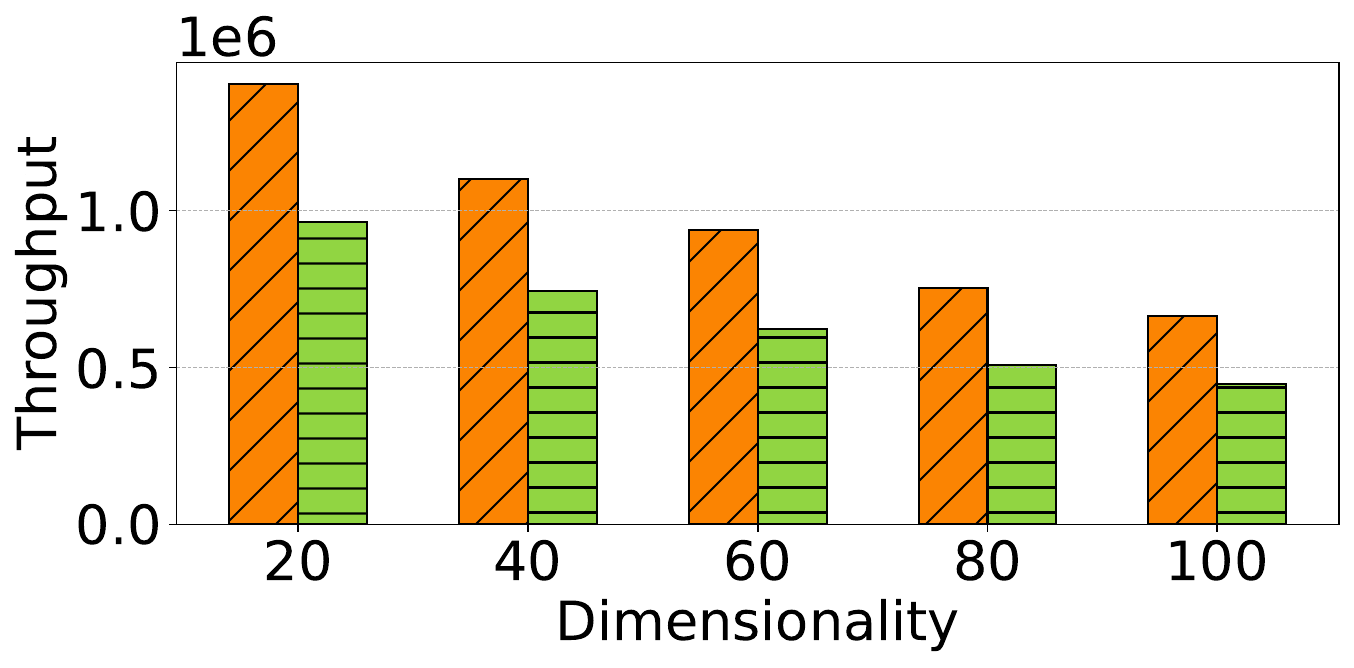}
        \caption{Efficiency}
        \label{fig:dim_throughput}
	\end{subfigure}     
        \caption{Performance Comparison on \dimension workload with varying dimensionality.}
        \label{fig:dim_comparison}	
\end{figure}

We now use three datasets \eds, \ods, and \dimension to evaluate the impact of varying workload characteristics.
Results are shown in Figure~\ref{fig:evolve_comparison} and Figure~\ref{fig:dim_comparison}.

\underline{First}, By evaluating \eds and \ods workloads, we show that \emph{\algo (Accuracy)} and \emph{\algo (Efficiency)} maintain their respective optimization targets even under frequent cluster or outlier evolution. This stability contrasts with the deteriorating performance observed in several existing \dsc algorithms, such as \clustream and \slkmeans, which struggle to adapt to evolving conditions. We attribute this resilience to \algo's dynamic composition capability. Unlike most existing algorithms, \algo continuously monitors stream characteristics to detect any changes, thereby enabling timely and accurate adaptations to the evolving data stream. This dynamic adaptability ensures superior clustering performance under varying conditions.

\underline{Second}, we next evaluate the impact of varying dimensions, utilizing the \dimension workload for this purpose. This workload comprises datasets with dimensions ranging from 20 to 100, as detailed in Table~\ref{tab:Workload}. Since existing baselines all predefined the dimensionality of the processed streaming data and unable to adjust it during the clustering process, we only conduct the comparison between the two variants of \algo. The outcomes of this evaluation are graphically represented in Figure~\ref{fig:dim_comparison}. Remarkably, \algo maintains a stable purity level of approximately 0.4 across datasets with diverse dimensions. While this purity level may not be high, its consistency is noteworthy, particularly when contending with the \emph{Curse of Dimensionality}. Operations integral to clustering—such as updating the summarizing data structure and pinpointing the appropriate cluster for data insertion—are intrinsically reliant on distance calculations involving the original high-dimensional data. The efficacy of these distance metrics tends to wane as the dimensionality escalates, thereby exacerbating the challenge of distinguishing between data points in high-dimensional spaces. Additionally, we discern a decrement in \algo's efficiency concomitant with an increase in the dataset's dimensionality. Our analysis ascertains that the computational complexity of several pivotal operations—including but not limited to the updating of the summarizing data structure and the selection of suitable clusters for data insertion—is directly influenced by the dimensionality of the workload. Consequently, a surge in dimensionality incurs a proportional rise in the computational time required for these operations, thereby attenuating the overall efficiency of the clustering process.

\subsection{Empirical Evaluation of Dynamic Composition Ability}
We conducted an empirical evaluation to demonstrate the effectiveness of \algo's dynamic composition abilities, including regular stream characteristics detection (Section~\ref{subsec:regular-detection}), automatic choice selection (Section~\ref{subsec:auto-selection}) and flexible algorithm migration (Section~\ref{subsec:migration}) respectively under evolving workload characteristics. 
To assess the role of the regular stream characteristics detection module, we estimated the general location where the stream characteristics evolve in the workload and checked whether the evolution was detected timely and accurately based on the output of \algo's regular detection function, as discussed in Algorithm~\ref{algo:auto_detection}. To evaluate the effectiveness of the automatic choice selection and flexible algorithm migration modules, we introduced two additional variants: \algo \textit{(Accuracy) without migration} and \algo \textit{(Efficiency) without selection}. By comparing the changes in both purity and throughput across these four variants of \algo, we identified the specific contributions of the automatic choice selection and flexible migration modules to the overall dynamic composition capability.

\begin{figure}[t]
	\centering
 \begin{minipage}{0.8\linewidth}
		\includegraphics[width=1\linewidth]{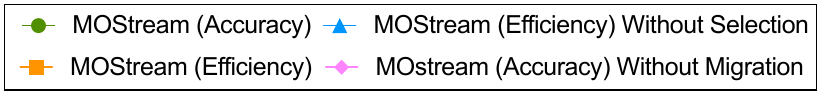}		
	\end{minipage}
 
	\begin{subfigure}{.235\textwidth}
        \centering
        \includegraphics[width=\linewidth]{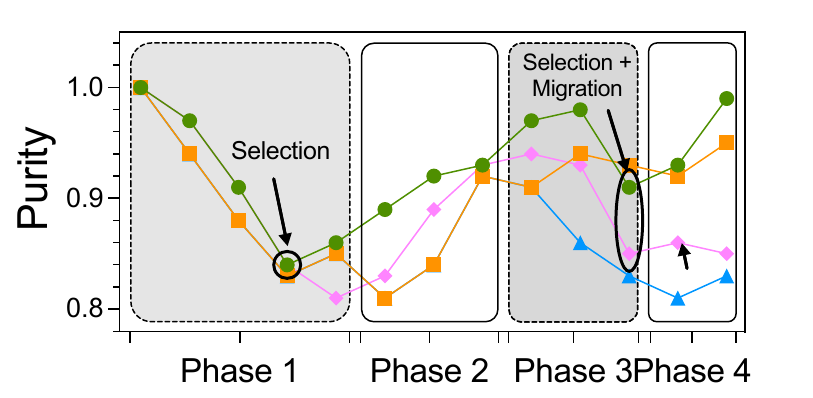}      
         \caption{Accuracy}
         \label{fig:dynamic_purity_kdd}
	\end{subfigure}       
  	\begin{subfigure}{.245\textwidth}
        \centering
        \includegraphics[width=\linewidth]{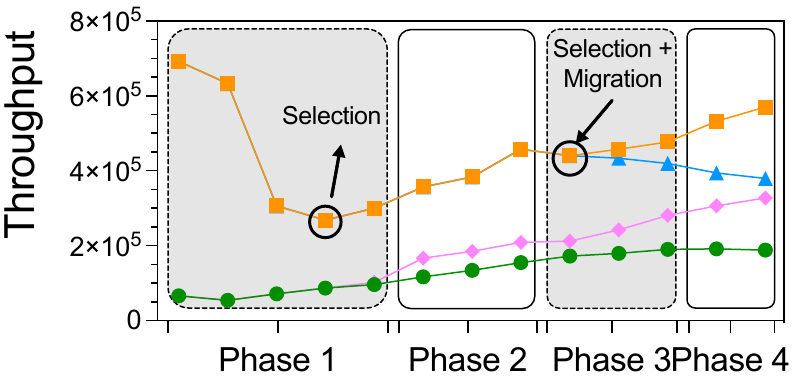}
        \caption{Efficiency}
        \label{fig:dynamic_throughput_kdd}
	\end{subfigure}     
    \caption{Detailed performance analysis of \algo's variants on \kdd. Measurements are taken at intervals of 350,000 data points with 4 phases in total.}
    \label{fig:dynamic_real_comparison}
\end{figure}

\subsubsection{Composition Effectiveness Study}
We commence by evaluating the performance of the four \algo variants on the real-world \kdd workload, characterized by a high frequency of outlier evolution and low but increasing frequency of cluster evolution. Measurements are taken at intervals of 350,000 data points with 4 phases in total, capturing both clustering outcomes and workload characteristic evolutions, as illustrated in Figure~\ref{fig:dynamic_real_comparison}. Notably, \algo (Accuracy) excels in purity, while \algo (Efficiency) demonstrates superior throughput, each aligning with their respective optimization objectives. Three key observations emerge from this analysis.

\underline{First}, a synchronized examination of \algo's clustering behavior and workload evolution (indicated by grey lines in the figure) reveals that while workload changes adversely affect clustering performance (evident in Phases 1 and 3), both \algo (Accuracy) and \algo (Efficiency) swiftly recover in Phases 2 and 4. For \algo (Accuracy), the initial use of the \damped window model leads to a rapid decline in both purity and throughput during Phase 1 due to its inability to effectively manage increasing outlier frequencies. However, the algorithm's regular stream characteristics detection module, as outlined in Algorithm~\ref{algo:auto_detection}, identifies this issue at the close of Phase 1, prompting a transition to the \landmark window model to better cope with outlier evolution. The subsequent improvement in purity during Phase 2 attests to the efficacy of this module. A similar recovery is observed in Phase 4, where the algorithm switches from \CFT to \MCs to adapt to increasing cluster evolution in Phase 3. For \algo (Efficiency), the algorithm opts to forgo outlier detection to minimize overhead, in line with its optimization target. Consequently, its performance deteriorates in Phases 1 and 2. However, upon entering Phase 3, characterized by high cluster evolution, the algorithm promptly detects this change and switches from \Grids to \DPT, resulting in improved purity and throughput in Phase 4, as depicted in Figure~\ref{fig:dynamic_real_comparison}.

\underline{Second}, apparently, both the purity and throughput drops significantly when canceling the usage of automatic design choice selection and switching, as shown by \algo \textit{(Efficiency) without selection}.  This indicates the limitation of the individual composition
On the contrary, applying the automatic design choice selection into the algorithm can make full use of the strengths of every design choices, leading to both better clustering accuracy and efficiency, as shown by \algo (Efficiency). \underline{Third}, the inclusion of migration in the algorithm improves accuracy at the expense of clustering speed. A comparative analysis of \algo (Accuracy) and \algo \textit{(Accuracy) without migration} reveals that while the former consistently outperforms the latter in purity, it lags in throughput. A detailed assessment of the overhead incurred by the three dynamic composition modules will be presented in the subsequent section.

\subsubsection{Analysis of Composition-Related Overhead}
\begin{figure}[t]
	\centering
	\begin{minipage}{1\linewidth}
        \centering
		\includegraphics[width=0.7\linewidth]{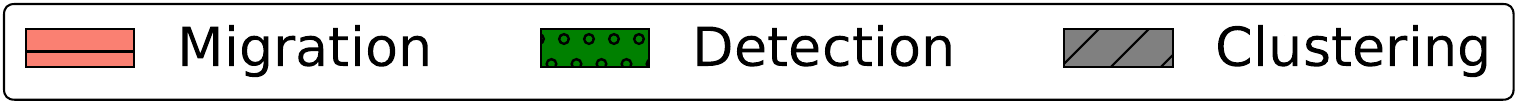}		
	\end{minipage}
	\begin{subfigure}{.24\textwidth}
        \centering
        \includegraphics[width=\linewidth]{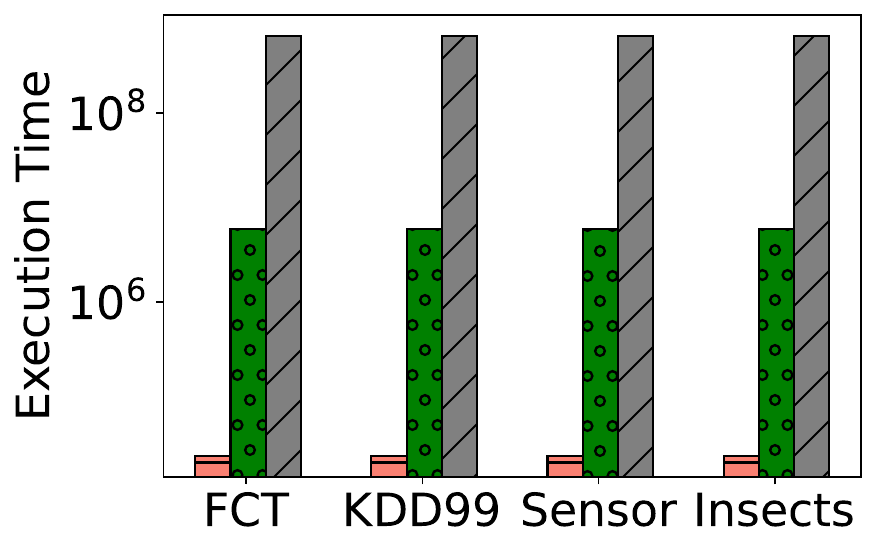}      
         \caption{\algo (Accuracy)}
         \label{fig:etb_purity}
	\end{subfigure}       
  	\begin{subfigure}{.24\textwidth}
        \centering
        \includegraphics[width=\linewidth]{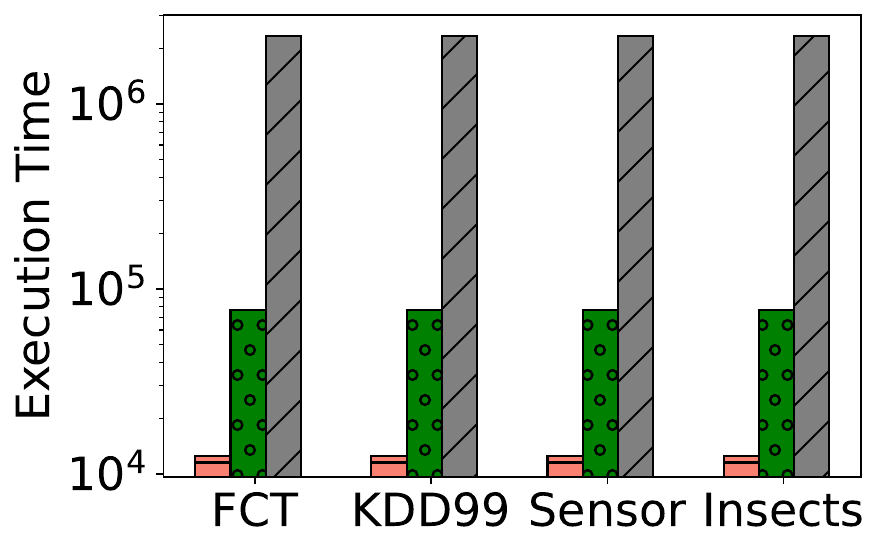}
        \caption{\algo (Efficiency)}
        \label{fig:etb_throughput}
	\end{subfigure}     
        \caption{Execution Time Break Down Analysis on Real-world Workloads of two \algo variants.}
        \label{fig:etb_analysis}	
\end{figure}

We proceed to examine the computational overhead associated with \algo's two pivotal composition procedures: detection and migration. This is juxtaposed against the time expended on clustering, contingent on the selected composition. As illustrated in Figure~\ref{fig:etb_analysis}, the time allocation for both detection and migration is relatively minimal for \algo (Accuracy) in comparison to the primary clustering task. This underscores the efficiency of these composition procedures.
In the case of \algo (Efficiency), the scenario is somewhat different. Given that this variant is optimized for speed, the clustering operation itself is more time-efficient. Consequently, the proportion of time spent on the detection procedure appears to be larger relative to \algo (Accuracy). However, it's crucial to note that \algo (Efficiency) omits the migration procedure altogether, as elaborated in Section~\ref{subsec:migration}. This strategic omission further enhances its efficiency, aligning it closely with its optimization objectives.
This analysis confirms that the overheads associated with dynamic composition in \algo are well-contained, thereby not compromising the algorithm's primary objectives of either accuracy or efficiency.

\subsection{Sensitivity Analysis of Parameters}
\begin{figure}[t]
	\centering
	\begin{subfigure}{.233\textwidth}
        \centering
        \includegraphics[width=\linewidth]{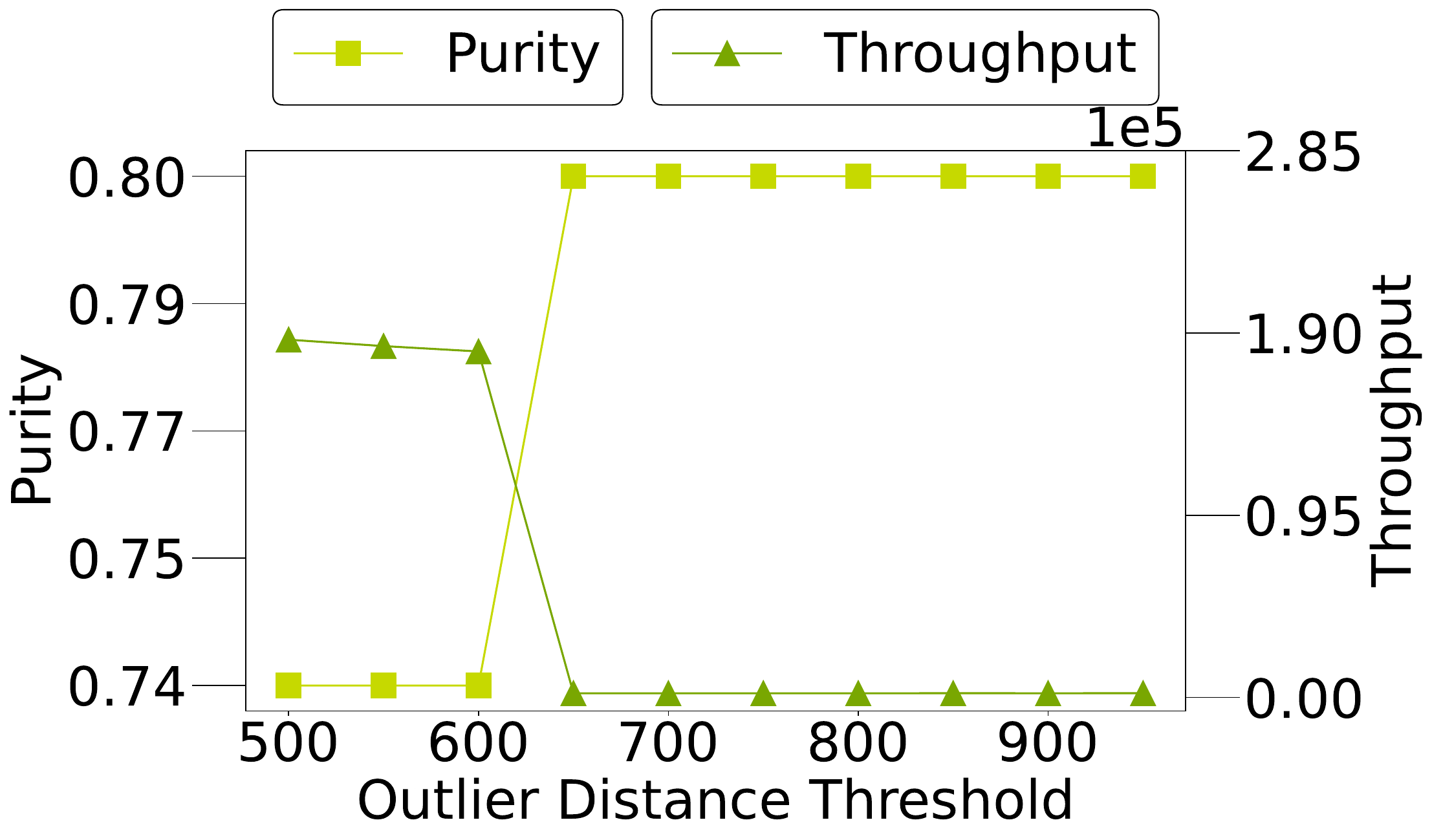}
         \label{fig:outlier_distance_purity}
	\end{subfigure}       
  	\begin{subfigure}{.244\textwidth}
        \centering
        \includegraphics[width=\linewidth]{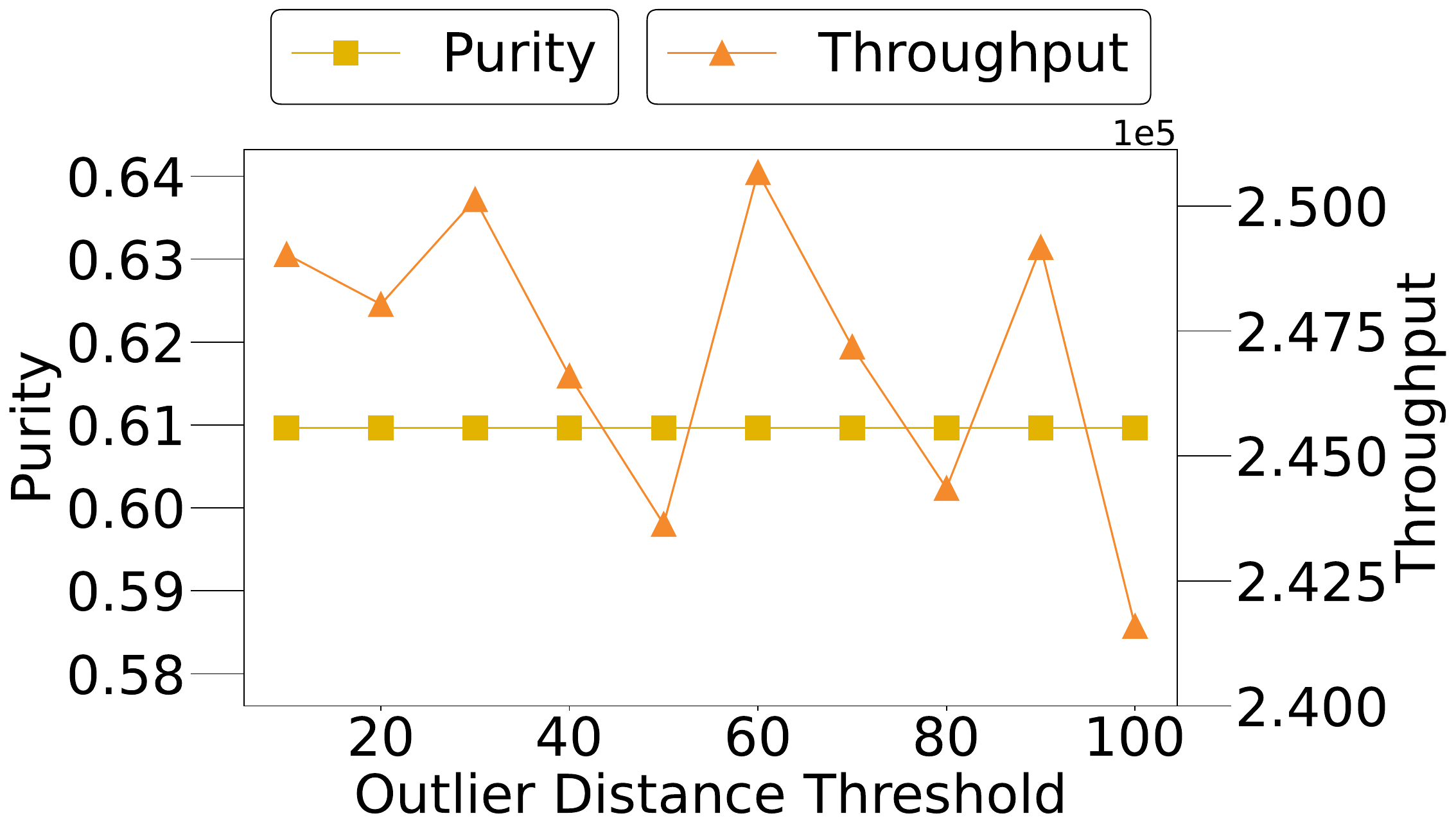}
        \label{fig:outlier_distance_throughput}
	\end{subfigure}
        \centering
	\begin{subfigure}{.238\textwidth}
        \centering
        \includegraphics[width=\linewidth]{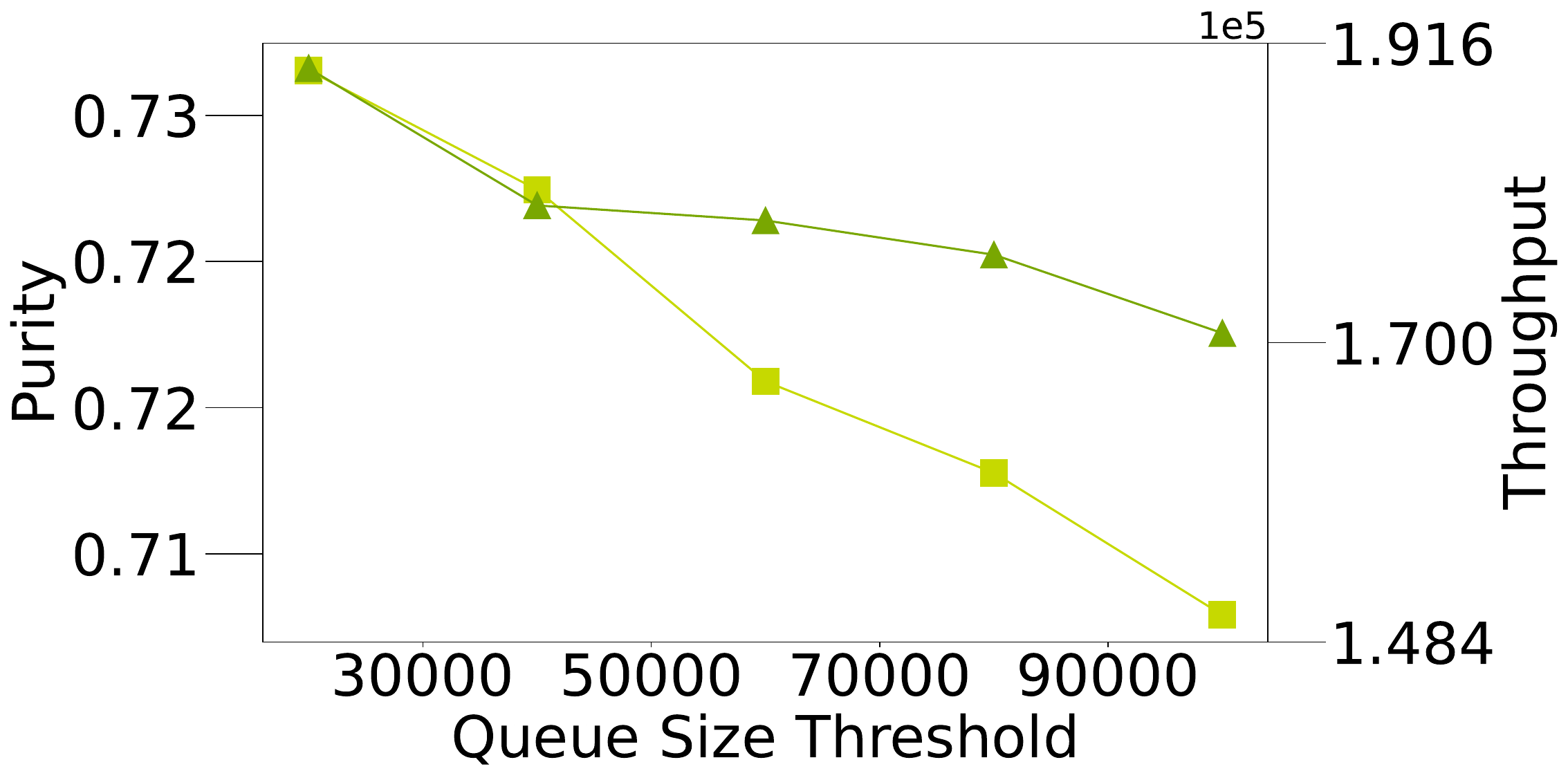}      
         \label{fig:queue_size_purity}
	\end{subfigure}       
  	\begin{subfigure}{.244\textwidth}
        \centering
        \includegraphics[width=\linewidth]{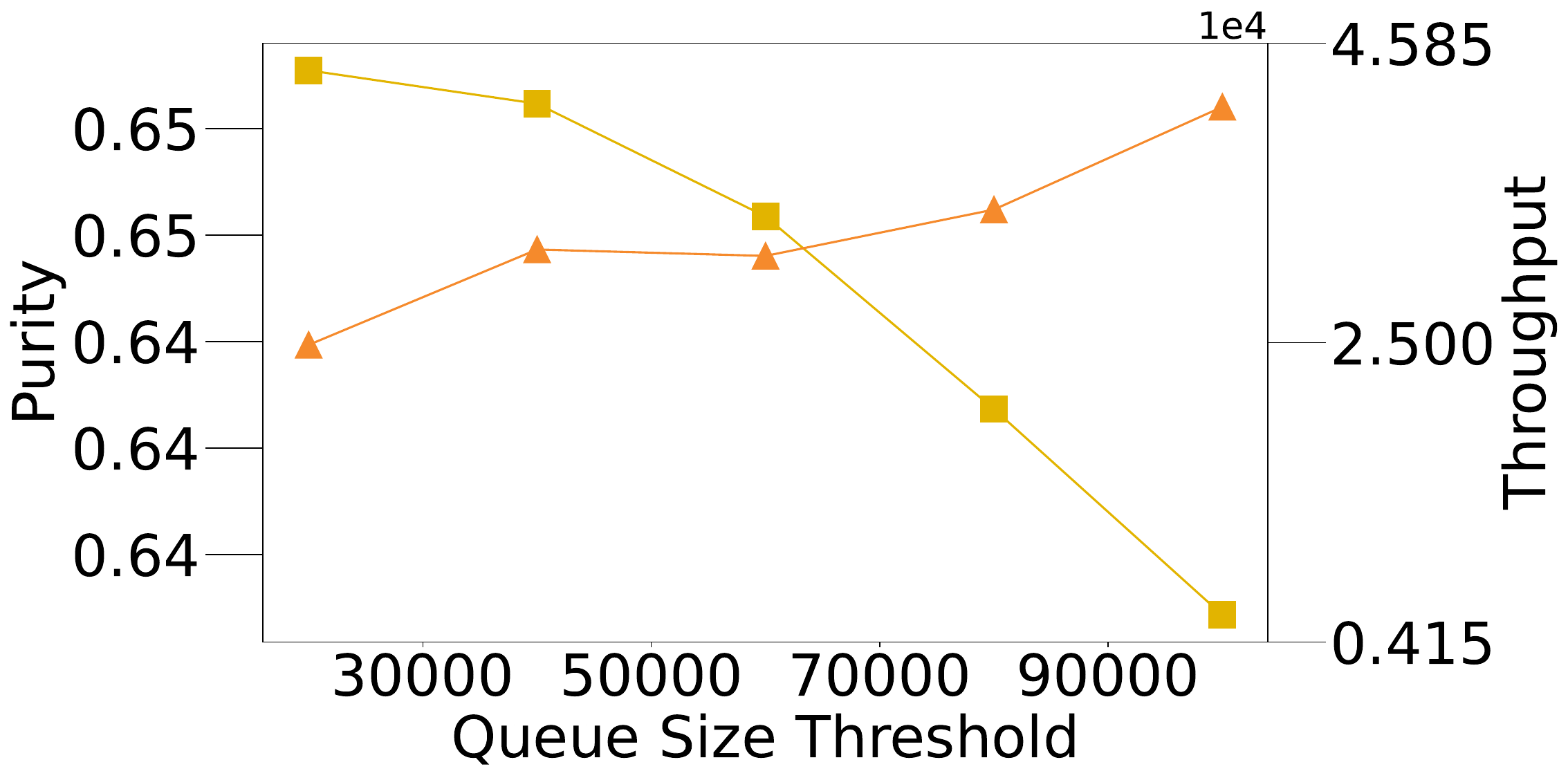}
        \label{fig:queue_size_throughput}
	\end{subfigure}     
        \centering
	\begin{subfigure}{.233\textwidth}
        \centering
        \includegraphics[width=\linewidth]{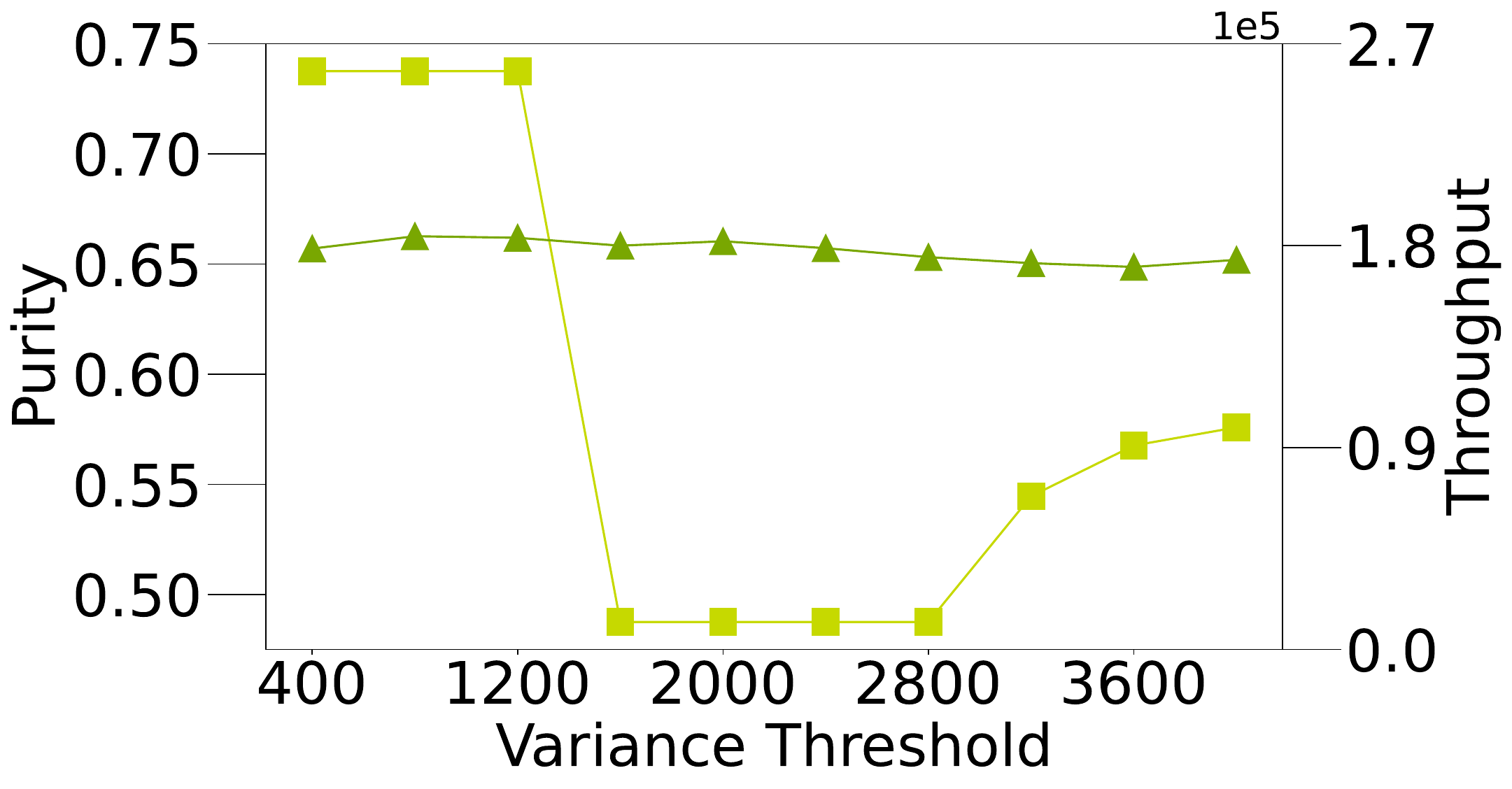}      
         \caption{\algo(Accuracy)}
         \label{fig:variance_purity}
	\end{subfigure}       
  	\begin{subfigure}{.244\textwidth}
        \centering
        \includegraphics[width=\linewidth]{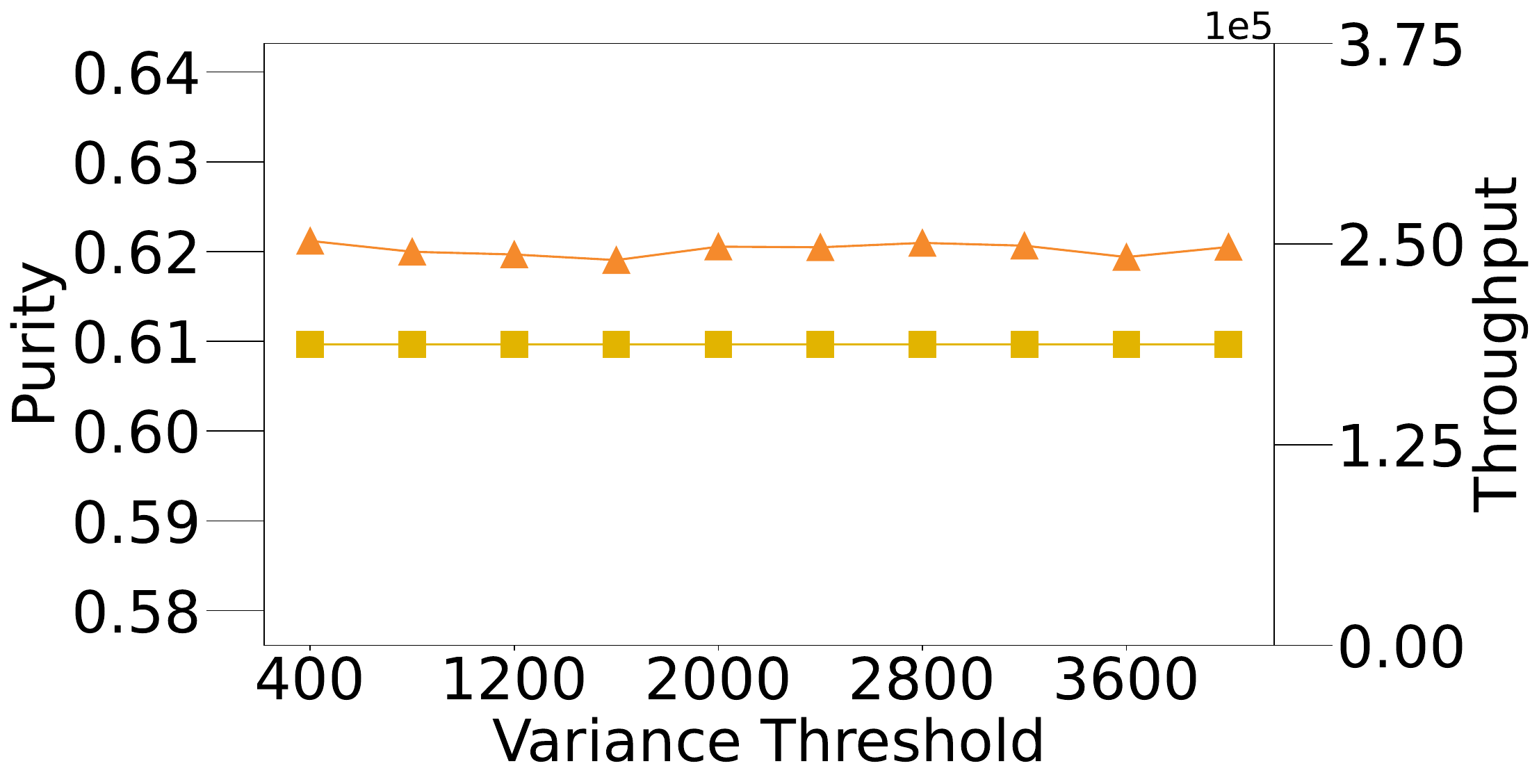}
        \caption{\algo(Efficiency)}
        \label{fig:variance_throughput}
	\end{subfigure}     
    \caption{Parameter Analysis of \algo variants on \fct.}
    \label{fig:parameter_analysis}
\end{figure}

We conducted a sensitivity analysis of \algo (Accuracy) and \algo (Efficiency) on \fct workload. As previously delineated, \algo (Accuracy) aims for elevated purity levels, while \algo (Efficiency) targets higher throughput. Both variants fulfill their respective optimization criteria. The results are shown in Figure~\ref{fig:parameter_analysis}.

1) \textbf{Outlier Distance Threshold ($\delta$)}: For each data point $x_t$ and its nearest cluster $C_t$, we compute the distance $D(x_t, C_t)$. If this distance surpasses the threshold $\delta$, $x_t$ is deemed an outlier. We experimented with $\delta$ values from 50 to 950 for \algo (Accuracy) and from 10 to 100 for \algo (Efficiency). Both variants maintain stable purity and throughput levels across this range. Notably, \algo (Accuracy) undergoes a window model transition from 'landmark' to 'damped' when the outlier distance threshold increases within a specific range, causing a sudden alteration in performance metrics, and leads to an increase in cluster size and purity. However, the throughput does not change, and this is due to the fact that both \landmark and \damped are time consuming for clustering. Conversely, \algo (Efficiency) remains unaffected as it does not consider the number of outliers as a performance metric.

2) \textbf{Queue Size Threshold}: We varied the queue size from 20,000 to 100,000. Both variants exhibit a decline in purity as the queue size increases, attributed to less frequent algorithmic adjustments. While the throughput for \algo (Accuracy) diminishes with an increasing queue size due to the growing complexity of its summarizing data structure, \algo (Efficiency) experiences a throughput increase. This is attributed to fewer algorithmic migrations and a transition to a more efficient data structure (\Grids) as the queue size enlarges.

3) \textbf{Variance Threshold}: \algo activates the \textit{characteristics.frequent evolution} flag when the calculated variance of sampled data exceeds a predefined threshold. We tested variance thresholds from 400 to 4,000. As the threshold rises, both purity and throughput for \algo (Accuracy) decline. This is due to the algorithm's assumption of infrequent evaluations at higher variance thresholds, leading to less frequent algorithmic migrations and increased computational overhead. \algo (Efficiency) remains relatively stable across varying variance thresholds. At a high variance threshold of 4,000, both variants achieve similar purity levels, but \algo (Efficiency) outperforms \algo (Accuracy) in throughput due to the absence of algorithmic migrations.

\compact

\section{Related Work} 
\label{sec:relatedwork} 
Data stream clustering algorithms have evolved significantly, yet many still lack the flexibility and self-optimization needed for diverse applications, resulting in suboptimal performance. Early work by Zhang et al.~\cite{zhang1997birch} introduced the Clustering Feature Tree (CFT) with a static structure, limiting adaptability. BIRCH~\cite{BIRCH:96} identified outliers based on density thresholds but did not adapt well to evolving data. Metwally et al.~\cite{metwally2005duplicate} proposed the landmark window model, while Zhou et al.~\cite{zhou2008tracking} and Borassi et al.~\cite{SL-KMeans:20} introduced variations of the sliding window model. These models improve adaptability but lack dynamic adjustment mechanisms. Aggarwal et al.~\cite{Clustream:03} extended CFT into microclusters (MCs) and introduced the online-offline clustering paradigm and the outlier timer, yet their fixed parameters limit adaptability. Cao et al.~\cite{cao2006density} and Chen et al.~\cite{DStream:2007} proposed the damped window model for temporal relevance, retaining all data but prioritizing recent information, though these models can be computationally expensive. Chen et al.~\cite{DStream:2007} also introduced a grid-based structure for efficiency. Wan et al.~\cite{DenStream:06} developed the outlier buffer, improving outlier management but adding computational overhead. Gong et al.~\cite{EDMStream:17} presented the Dependency Tree (DPT), balancing efficiency and accuracy but struggling with complex features. Farnstrom et al.~\cite{farnstrom2000scalability} and Bradley et al.~\cite{bradley2002scaling} proposed scalable and streaming k-means variants for large-scale data, yet these methods lack mechanisms for dynamic adjustment to data stream variations. These limitations highlight the need for a more flexible and adaptive approach. Our proposed algorithm, \algo, addresses these gaps by offering a modular and self-optimizing framework that dynamically balances clustering accuracy and efficiency, outperforming existing methods in varied data stream scenarios.

\compact
\section{Conclusion}
\label{sec:conclusion}
This paper introduced \algo, a \dsc algorithm that autonomously selects optimal configurations based on user-defined objectives and input data stream characteristics. Empirical evaluations demonstrate that \algo surpasses state-of-the-art algorithms in both accuracy and efficiency by optimally selecting combinations of four key design aspects of \dsc algorithms. Future research will focus on extending \algo's capabilities to accommodate high-dimensional data streams, enhancing the automated configuration selection process, and integrating advanced machine learning techniques into the framework of \algo.

\bibliographystyle{IEEEtran}
\bibliography{References}

\end{document}